\documentclass[12pt]{JHEP3}
\usepackage{graphicx}
\usepackage{dcolumn}  
\usepackage{bm}    
\usepackage{amssymb} 
\usepackage{amsmath,bm}
\usepackage{amsfonts}    
\usepackage{slashed}  
\usepackage[mathscr]{euscript}
\title{Supersymmetry of classical solutions  in Chern-Simons  higher spin supergravity}
\author{
Shouvik Datta,  Justin R. David   \\
 Centre for High Energy Physics,
Indian Institute of Science,\\ C.V. Raman Avenue, Bangalore 560012, India. \\
\email{shouvik, justin@cts.iisc.ernet.in}\\
}
\abstract{ 
We construct and study  classical solutions in Chern-Simons supergravity based on the 
superalgebra $sl(N|N-1)$. 
The algebra for the  $N=3$ case 
is written down explicitly using the fact that it  
arises as  the global part of the 
super conformal   ${\cal W}_3$  superalgebra. 
For this case we construct new classical 
solutions and study their supersymmetry. 
Using the algebra we write down the Killing spinor equations and explicitly 
construct the Killing spinor for conical defects and black holes  in this theory. 
We show that  for the  general $sl(N|N-1)$ theory  the 
condition for the periodicity of the Killing spinor can be written in 
terms of the products of the odd roots of the super algebra
and the eigenvalues of the holonomy matrix of the background.  Thus the supersymmetry 
of a given background can be stated in terms of gauge invariant and 
well defined physical observables of the Chern-Simons theory.  
We then show that  for 
$N\geq 4$,  the $sl(N|N-1)$ theory admits  
smooth supersymmetric conical defects. 
}

\begin{document}
\def\G{\Gamma}
\def\lb{\left(}
\def\rb{\right)}
\def\k{\kappa}
\def\nn{\nonumber}
\def\pU{U^+}
\def\mU{U^-}
\def\pmU{U^\pm}
\def\pG{G^+}
\def\mG{G^-}
\def\pmG{G^\pm}
\def\nn{\nonumber}

\section{Introduction}

Consistent theories of  interacting higher spin fields  constructed by Vasiliev 
\cite{Vasiliev:2003ev}  have been the focus of many recent works.
For a review of higher spin theories see   \cite{Bekaert:2005vh}
These theories are interesting
from the perspective of the AdS/CFT since they are examples
of gravitational  backgrounds 
in which one does not need to deal with the  entire spectrum of massive string excitations
but only with  infinite set of higher spin  fields. 
Higher spin theories on $AdS_4$ have been proposed as dual descriptions of 
vector like field theories 
\cite{Klebanov:2002ja,Giombi:2009wh, Giombi:2010vg}, 
sub-sectors of free Yang-Mills theories 
\cite{ Sundborg:2000wp,Mikhailov:2002bp,Sezgin:2002rt,Douglas:2010rc},  and very recently
argued to be  duals of  certain ABJ models \cite{Chang:2012kt}. 

Higher spin theories in three space time dimensions are particularly tractable since in this 
situation the Vasiliev like theories can be formulated in terms of a Chern-Simons theory
\cite{Blencowe:1988gj}. 
Furthermore in three dimensions,  it is not necessary to consider an infinite number of
higher spin fields to obtain consistent interactions. It is possible to work with a finite
set of higher spin fields. 
Vasiliev like theories in 3 dimensions  coupled to a massive complex scalar have been 
proposed to be holographic duals to ${\cal W}_N$ minimal models based on the 
coset \cite{Gaberdiel:2010pz}
\begin{equation}
\frac{SU(N)_k \otimes SU(N)_1}{SU(N)_{k+1}}. 
\end{equation}
This duality is a new example for the $AdS_3/CFT_2$ correspondence and various 
checks of the proposal include matching of the symmetries, comparison of the
one loop partition function and the  three point correlators.
For a comprehensive list of references please  see \cite{Gaberdiel:2012uj}.   
A supersymmetric extension of the higher spin/minimal model duality has been 
proposed in \cite{Creutzig:2011fe}. This duality has also been checked by comparison of the 
symmetries and the partition function \cite{Candu:2012jq,Candu:2012tr}.  
Chern-Simons theories based on
super-extended higher spin super algebras have been considered in \cite{Henneaux:2012ny} and 
their asymptotic algebras have been shown to agree with the corresponding
super conformal ${\cal W}_\infty$ algebra.  

Studying classical solutions in Chern-Simons theories based on  a higher spin 
group provides more more insights to holography in three dimensions. 
The higher spin black holes found in \cite{Gutperle:2011kf} and the conical defect solutions 
\cite{Castro:2011iw} have proved to be useful to study aspects of the holographic renormalization 
group and the nature of singularities in higher spin gravity \cite{Ammon:2011nk,Castro:2011fm}.
In fact the higher spin black hole studied in \cite{Gutperle:2011kf} is dual to a 
renormalization  group flow between two CFT's. 
Smooth conical defects have been argued to be dual to the primaries in the $\mathcal{W}_N$ minimal 
model after an appropriate analytic continuation \cite{Castro:2011iw}.

Motivated by these 
developments   in bosonic higher spin theories 
we study  and construct new classical solutions in Chern-Simons theories based on 
the $sl(N|N-1)$ super algebra. 
Any classical  solution of the bosonic theory can be embedded as a
 solution in the supersymmetric theory. In addition to these solutions,  supersymmetric theories
 also admit solutions in which additional fields required for the supersymmetric completion 
 are turned on. In this paper we find such  solutions in the Chern-Simons theory 
 based on  $sl(3|2)$ super algebra. These solutions  are conical defects and black holes, 
 which  have fields valued in $sl(2)$ and the $u(1)$ part of the connection 
 in addition to the $sl(3)$. 
 The main motivation to construct the  solutions in this paper is to  study
 their supersymmetry. 
 The study of supersymmetry in higher spin theories is a new subject and 
 as far as we are aware there are no general results for  when a classical 
 solution is supersymmetric in higher spin theories. 
  An early study of supersymmetry  of a 
 black hole solution in a higher spin theory  on $AdS_4$ is  \cite{Didenko:2009td}. 
 The Killing spinor equations for this case are quite involved and difficult 
 to solve. We will see  that Killing spinors   for solutions in supersymmetric higher 
 spin Chern-Simons theories are  considerably  easier to obtain. 
 In fact we will obtain a general condition for when a classical solution is supersymmetric. 
 This condition can be stated invariantly in terms of the eigenvalues of the
 holonomy  of the classical 
 solution. This is important since the  Chern-Simons action is independent
 of the metric on the manifold and the eigenvalues of the holonomy are 
 the only gauge independent well defined physical observables. 
 
 Our working example will be the 
algebra $sl(3|2)$ which is the global part of the   ${\cal W}_3$ super algebra
in the large central charge limit. All these theories have two $U(1)$ gauge fields
corresponding to the $R$ symmetry of the dual conformal field theory. 
The supersymmetric conditions in the Chern-Simons theory based on 
supergroups which contained spins $\leq 2$ were earlier analyzed in 
\cite{Coussaert:1993jp,Izquierdo:1994jz,David:1999zb,Balasubramanian:2000rt,Maldacena:2000dr}. 
Once the background flat connection is given,
the Killing spinor equations are particularly easy to write 
 for a Chern-Simons theory based on   any supergroup. 
 The Killing spinor   equation is 
 just a covariant derivative with the flat connection as the background. 
 Thus the supergroup structure is sufficient to write down the Killing 
 spinor equation.  By studying various solutions we arrive at the observation
 that the solution admits 
 a periodic Killing spinor if  the combined $U(1)$ part of the holonomy together with the 
 holonomy of the rest of the connection 
  around the angular direction  in $AdS_3$ is trivial. 
This observation enables us to state the condition on the 
periodicity of the Killing spinor in terms of the  odd roots of the
$sl(N|N-1)$ super algebra and the eigenvalues of the holonomy matrix.  
This is one of  the key results of this work.   The reader can directly 
turn to section \ref{spinper} and the refer to equation (\ref{periodf}) for this result. 
Using this condition for the periodicity of the Killing spinor we show that for 
$N\geq 4$ the $sl(N|N-1)$ theory admits smooth supersymmetric 
conical defects. 
These solutions should  play a crucial role in obtaining the duals of the chiral primaries 
in the supersymmetric minimal ${\cal W}_N$ model proposed in \cite{Creutzig:2011fe}.

The organization of this paper is as follows: In the next section we review some 
generalities of higher spin $AdS_3$ supergravity and write down the Killing 
spinor equation for any Chern-Simons theory based on a given super group. 
We then provide the details of the commutation relations for the
$sl(3|2)$ super algebra. We derive them by considering the global part
and the large central charge limit of the super ${\cal W}_3$ conformal 
algebra written down in \cite{Romans:1991wi}. 
In section 3 we study the supersymmetry of various classical 
solutions in the Chern-Simons  theory based on the $sl(3|2)$ super algebra. 
These include the BTZ black hole, the black hole with higher spin field.
They also include a new black hole solution,
this background has   fields  valued in  the $sl(2)$ required 
for the supersymmetric completion of the bosonic $sl(3)$ turned on. 
We then study the supersymmetry of conical defects in these theories. 
Again these defects also include those with fields in the $sl(2)$ turned on. 
A summary of  the supersymmetric conditions for these backgrounds is provided 
in table 1 of section \ref{table}.
In section 4 we show that the periodicity requirement of the Killing spinor in the angular 
direction can be  cast in terms of the holonomies  of the background flat 
connection. We show that  the supersymmetric
conditions of any background  can be written in terms of  products of 
the odd roots of the super algebra with eigenvalues of the holonomy matrix 
of the background. 
We then use this result to show that for $N\geq 4$, the $sl(N|N-1)$ theory admits
smooth supersymmetric conical defects. 
Section 5 contains the conclusions and a discussion of the results. 

{\bf Note added:} After completion of this work, we received  \cite{Tan:2012xi} which overlaps 
with  some portions of this paper.

\section{Chern-Simons higher spin supergravity}

It is well known that pure gravity in $AdS_3$ can be written in terms of 
difference of two Chern-Simons actions based on the algebra  $sl(2, R)$
\cite{Witten:1988hc}.  Similarly supersymmetric extensions of pure gravity containing spins
$\leq 2$ can be written as a Chern-Simons action based on 
supersymmetric extensions of $sl(2,R)$ \cite{Achucarro:1987vz}. 
Since higher spin theories containing only bosonic fields are based on the 
the $sl(N, R)$ with $N>2$ \cite{Henneaux:2010xg,Campoleoni:2010zq}, it is natural to look for supersymmetric extensions
of  the $sl(N, R)$ algebra  to construct consistent interacting higher spin theories
in $AdS_3$ containing fermions. 
Given any such super algebra  ${\cal G}$ 
the parity invariant Chern-Simons action is given by 
\begin{equation}
S = \frac{k}{2\pi} \int\left[  {\rm str} \left(   \Gamma d\Gamma + \frac{2}{3} \Gamma^3 \right ) 
-  {\rm str} \left( \tilde\Gamma \tilde d\Gamma + \frac{2}{3} \tilde \Gamma^3 \right) \right]. 
\end{equation}
Here $\Gamma, \tilde \Gamma $  are the  1-forms which take values in ${\cal G}$ 
and ${\rm str}$ refers to the super-trace over the respective
algebras. The  
integral is over  the 3 dimension space time. 
The equations of motion of this action are  the following flatness conditions
\begin{equation} \label{eom}
d\Gamma + \Gamma\wedge \Gamma =0, 
\qquad
d\tilde\Gamma + \tilde\Gamma\wedge\tilde\Gamma =0. 
\end{equation}
To obtain the equations of motion in component form, one needs to 
expand  $\Gamma, \tilde\Gamma$ in terms of the generators of the super algebra.
The coefficients of this expansion  are the fields of the theory, this is then  
substituted in the equations given in (\ref{eom}) to obtain the equations of motion 
in the component form. Thus to write down the equations of motion it is sufficient
to know the structure  constants of the algebra. 

\vspace{.6cm} \noindent
{\bf The generalized  Killing spinor equations }
\vspace{.3cm} \\
It is also easy to write down the Killing spinor equations. Let the  bosonic generators of 
the algebra be denoted by $T_a$ and the corresponding bosonic fields by $A^a$. 
Similarly let the fermionic generators be
  $G_i$. 
  Consider a 
bosonic solution to the equations of motion. Then 
one has the following equation
\begin{equation}\label{beom}
d(A^a T_a) + ( A^a T_a )  \wedge ( A^b  T_b) =0,  
\end{equation}
where the bosonic fields $A^a$ are 1-forms.  
The Killing spinor equation is essentially the equation that demands that the 
background $A^a$ is invariant under fermionic gauge transformations. 
Let $\epsilon^i$ be the parameters of this transformation, then the  
equation for the Killing spinor is given by 
\begin{equation}\label{kill}
\delta \psi \equiv \partial_\mu  \epsilon^i  G_i + A^a_\mu \epsilon^i [ T_a ,  G_i] =0 .   
\end{equation}
This is essentially the equation demanding that the  covariant derivative in 
presence of the bosonic background $A^a_\mu$ vanishes. 
The solutions $\epsilon^i$ are the Killing spinors. 
It is clear that the variation \(\delta \psi\) is a fermionic symmetry of the 
Lagrangian since it is just a gauge transformation. 
By demanding $\delta \psi =0$ we are looking for 
general variations with parameters involving  fermions
which leaves the background invariant.  In general the fermionic fields  $\epsilon^i$
 can contain fermions with spins $s\geq 1/2$. 
This is the generalized notion 
of the Killing spinor in the higher spin theory. 
It is important to note that flatness conditions in (\ref{beom}) are the integrability 
constraints of the Killing spinor equation (\ref{kill}). Thus given that a
bosonic background satisfies the equations of motion,  solutions to the Killing 
spinor equations are guaranteed to exist. However we must also the
impose the condition that the Killing spinors are periodic with respect to the 
angular co-ordinate in $AdS_3$. This then decides the condition 
whether a given background is supersymmetric. 

The class of super algebras we will be interested in belongs to
$sl(N|N-1)$.  We will also examine the supersymmetry in  one copy of the 
$sl(N|N-1) \oplus sl(N|N-1)$ Chern-Simons theory. 
However the central conclusion regarding 
the supersymmetry  of a given background in terms of the 
eigenvalues of the holonomy of the background  drawn at the end our analysis
applies to any super algebra. 
An appropriate basis to discuss the $sl(N|N-1)$ algebra is the explicit matrix representation 
of the algebra given in section 61 of \cite{Frappat:1996pb}. This is in the Cartan-Weyl basis 
which is suitable for the general analysis of 
the Killing spinor and the supersymmetric conditions. 
 We will explicitly study the case of $sl(3|2)$. The bosonic part of this algebra 
is given by  $sl(3) \oplus sl(2) \oplus u(1)$. This algebra contains the super group $sl(2|1)$ on which
$(2, 2)$ supergravity in $AdS_3$ is based.

\subsection{The $sl(3|2)$ superalgebra}

In this section we  write down  the commutation relations of $sl(3|2)$. 
We obtain this by taking the large central charge and the global part of the 
${\cal N}=2$ super ${\cal W}_3$ algebra written down by \cite{Romans:1991wi}. 
This provides evidence  
evidence that the boundary theory of Chern-Simons gravity based on 
$sl(3|2)$ is a super conformal theory  with ${\cal N}=2$ super ${\cal W}_3$ 
symmetry. 

${\cal N}=2$ super ${\cal W}_3$ algebra contains 
generators with  $J, G^{\pm}, L$ with spin $1, 3/2$ and $2$ respectively. 
These generators obey the ${\cal N}=(2,2)$ super conformal algebra 
among themselves. $J$ is the generator of the R-symmetry, $G^\pm$ are the supersymmetry
generators and $L$ is the stress tensor. 
In addition to this there is also the 
generators $V, U, W$ with spin $2, 5/2, 3$ respectively.  $W$ generates the 
super conformal ${\cal W}_3$ symmetry. 
Taking the large central charge limit and the global part of the commutation 
relations of ${\cal N}=2$ super conformal ${\cal W}_3$ we obtain the following
algebra for the bosonic generators:
\begin{eqnarray}\label{bos}
[J,J]&=&0, \qquad 
[L_m,L_n]=(m-n)L_{m+n},   \\ \nonumber
[V_m,V_n]&=& (m-n)(L_{m+n} + \k V_{m+n}), \\ \nonumber
[W_m,W_n]&=&\tfrac{1}{4} (m-n)(2m^2+2n^2-mn-8) ( L_{m+n} + \tfrac{\k}{5}  V_{m+n} ),  
\\ \nonumber 
 [ J, L_n ]&=&0,  \qquad
[J,V_n]=0,  \qquad 
[J, W_n]=0 ,   \\ \nonumber
[L_m,V_n]&=&(m-n)V_{m+n},  \qquad 
[L_m,W_n]=(2m-n)W_{m+n},  \\ \nonumber
[V_m,W_n] &=&\tfrac{\k}{5}(2m-n)W_{m+n}.  
\end{eqnarray}
Here the subscripts $m,n$ on the generators $L$ run from $-1, 0, 1$ while the 
subscripts on the generators $W$ run from $-2, -1, 0, 1, 2$. 
The commutation relations between bosonic and fermionic generators are given by 
\begin{eqnarray} \label{bosf}
[L_m,\pmG_r]&=&  (\tfrac{1}{2}m - r) \pmG_{m+n},  \qquad 
[J, G^\pm_r] = \pm G^\pm _{r},   \\ \nonumber
[L_m , U^\pm _r] &=&  (\tfrac{3}{2}m-r)U^\pm_{m+r},   \qquad
[J , U^\pm _r] = \pm U^\pm_{r},   \\ \nonumber
[V_m ,G^\pm_r ] &=& \pm U^\pm_{r+m},  \qquad
[G^\pm_r, W_m]=(2r -\tfrac{1}{2}m)U^\pm_{r+m},  \\ \nonumber
[V_m, U^+_r] &=& \tfrac{2}{5}\k ( \tfrac{3}{2} m -r ) \pU_{m+r} + \tfrac{1}{4} (3m^2 - 2mr + r^2 -\tfrac{9}{4} ) \pG_{m+r},  \\ \nonumber
[V_m, U^-_r] &=& -\tfrac{2}{5}\k^* ( \tfrac{3}{2} m -r ) \mU_{m+r} - \tfrac{1}{4} (3m^2 - 2mr + r^2 -\tfrac{9}{4} ) \mG_{m+r},  \\ \nonumber
[U^+_r,W_m]&=& \tfrac{\k}{10} (2r^2-2rm+m^2-\tfrac{5}{2})U^+_{r+m}   \\  \nonumber
& & \qquad + \tfrac{1}{8} ( 4r^3-3r^2 m +2rm^2-m^3 - 9r +\tfrac{19}{4}m ) G^+ _{r+m}, 
 \\ \nonumber
[U^-_r,W_m]&=& \tfrac{\k^*}{10} (2r^2-2rm+m^2-\tfrac{5}{2})U^-_{r+m} \\ \nonumber
& & \qquad + \tfrac{1}{8} ( 4r^3-3r^2 m +2rm^2-m^3 - 9r +\tfrac{19}{4}m ) G^- _{r+m}. 
\end{eqnarray}
Here the subscripts $r, s$ on $G^\pm$ run from $-1/2, 1/2$ while the subscripts
on the generators $U^\pm$ run from $-3/2, -1/2, 1/2, 3/2$. 
Finally the anti-commutation rules between the fermionic generators are given by 
\begin{eqnarray}
\{  G^\pm_r, G^\mp_s \} &=&  2L_{r+s} \pm (r-s) J, \qquad  
\{  G^\pm_r , G^\pm_s  \} =0,  \\ \nonumber
\{   G^\pm_r , U^\mp _s      \}  &=&  2W_{r+s} \pm (3r-s)V_{r+s}, \qquad 
\{ G^\pm _r , U_s^\pm \} = 0 , \\ \nonumber
\{ \pU_r , \mU_s \} &=&  -\tfrac{2}{5}\k (r-s) W_{r+s}+( 3s^2-4rs+3r^2-\tfrac{9}{2} )( \tfrac{1}{2}L_{r+s}+\tfrac{\k}{5} V_{r+s}   )  \\ \nonumber
& & \qquad \ + \tfrac{1}{4} (r-s) (  r^2 +s^2 -\tfrac{5}{2} ) J_{r+s},  \\ \nonumber
\{  \pmU_r , \pmU_s  \} &=& 0 . 
\end{eqnarray}
On taking large central charge limit,  the non-linear 
terms in the super ${\cal W}_3$ algebra drop off  and  we obtain 
 $\k=\pm (5/2)i$.
We have verified that all the Jacobi identities of this algebra are satisfied using  the
Quantum add-on for Mathematica \cite{gomez}. 

\def\tr{\text{Tr}}

To see that the bosonic part of the algebra given in (\ref{bos}) is given by the 
direct sum   $sl(3) \oplus sl(2) \oplus u(1)$, we define the following linear combinations
of generators
\begin{equation}\label{decouple-1}
T^+_m = -\frac{1}{3}(L_m + 2i V_m) \, \qquad T^-_m = \frac{1}{3}(4 L_	m + 2i V_m). 
\end{equation}
Substitituting these redefintions in (\ref{bos}) we  obtain
\begin{equation}
[T^+_m,T^-_n]=0 \ , \qquad [T^+_m,W_n]=0 , 
\end{equation}
and the can show that the generators  $T^+_m$ obey the $sl(2)$ algebra while the 
generators $ T^-_n, W_m$ obey the  commutation relations of the $sl(3)$ algebra 
given by 
\begin{eqnarray}
 [T^-_m , T^-_n] &=& ( m-n) T^-_{m+n}, \qquad 
[T^-_m, W_n] = ( 2m -n) W_{m+n}, \\ \nonumber
[W_m , W_n] &=& \frac{3}{16} ( m-n) ( 2m^2 + 2n^2 -mn-8 ) T_{m+n}^-.
\end{eqnarray}
Note the comparing the $sl(3)$ algebra given in equation (A.2) of 
\cite{Gutperle:2011kf} we see that 
the parameter $\sigma $ defined in those equations is  equal to $(3/4)^2$. 
Now that we have the explicit $sl(3|2)$ algebra we can proceed to obtain 
solutions to the equations of motion  and study their supersymmetry.  The traces of the 
product of any two of the $sl(3)$ generators is the same as that of equation (A.3) of 
\cite{Gutperle:2011kf} with $\sigma = (3/4)^2$, while for the $sl(2)$ we use 
the representation in terms of the Pauli matrices.  

\def\cL{\mathcal{L}}
\def\cW{\mathcal{W}}
\def\cD{\mathcal{D}}
\def\e{\epsilon}
\def\pd{\partial}
\def\be{\tilde{\e}}
\def\l{\lambda}
\def\tl{\tilde{\lambda}}
\def\tp{T^+}
\def\tm{T^-}
\def\R{{\cal R}}
\def\M{\mathbb{M}}

\section{Supersymmetry of classical solutions}

 We begin this section by describing the general strategy we adopt to find the 
Killing spinors for the various backgrounds considered in this paper.
We reduce the  Killing spinor equation to  a set of 
ordinary first order equations with constant coefficients  which can then be easily 
solved. In section 3.2 we  construct the general higher spin conical defects in 
the $sl(3|2)$ theory. These solutions in general have fields in  the  
$sl(3)\oplus sl(2) \oplus u(1)$ directions. We then solve the Killing spinor equations 
and determine the supersymmetric conditions for the supercharges with 
 $u(1)$ charge in one copy   of
$sl(3|2)$ in the Chern-Simons theory. This analysis can be generalized for 
the remaining charges. We also determine  the special 
values in the parameter space of  conical defects which reduce to $AdS_3$. 
In section 3.3 we study the supersymmetry of black holes in this theory. 
This includes the usual BTZ black hole embedded in $sl(2)$, the 
higher spin black hole of \cite{Gutperle:2011kf} embedded in $sl(3)$ along with the $u(1)$ turned on. 
We also construct a new black hole solution which  has charges in 
$sl(3)\oplus sl(2) \oplus u(1)$ and study its supersymmetry. 
The list of all the solutions studied  and the corresponding supersymmetry conditions 
is given in table 1. 

\subsection{General strategy to obtain the Killing spinors}

The gauge connections in the $sl(3|2)$ theory  which will be of interest 
 in this paper has the following generic form 
\begin{align}\label{type-con}
A=& \left( \sum_{m=-1}^1 ( t_m e^{m\rho} \tm_m + s_m e^{m\rho} \tp_m )	 + 
 \sum_{m=-2}^2 ( w_m e^{m\rho} W_m )	+ \xi	J	\right) dx^+ \nn \\ &  \qquad  \hspace{7cm} -\xi J dx^- + (\tp_0+\tm_0) d\rho. 
\end{align}
Here $x^\pm = t \pm \phi$ and $\rho, t, \phi$ are the radial, time and the angular co-ordinates
of the three  dimensional spaces we consider. The connection in (\ref{type-con}) obeys the flatness condition. The general form of the connection can be conveniently written as $A=a_{\mu}^ne^{(n)\rho}T_n $. $T_n$ being a bosonic generator of the superalgebra. Negative weights of the generators 
with respect to $\L_0 = tp_0 + \tm_0 $ 
 appear in the exponential factors. For example, we have terms like $w_+^{(2)}e^{2\rho}W_{2}$ and $t_+^{(-1)}e^{-\rho}\tp_{-1}$.

The equation for Killing spinor
is given by 
\begin{equation}\label{kill3}
(\pd_\mu \epsilon^r)G_r + \epsilon^a A_\mu ^b [T_b,G_a]=0,  
\end{equation}
where $ [T_b,G_a]$ is some linear combination of the fermionic generators which we can write as 
$f_{bac} G_c$.  Here $f_{bac}$  are the structure constants of the superalgebra and 
 $b$ is a bosonic index while $a$ and $c$ are fermionic ones. 
 Substituting for the commutation relation in (\ref{kill3}) we obtain the  following equation
\begin{equation}
(\pd_\mu \epsilon^r)G_r + \epsilon^a A_\mu ^b f_{bac} G_c  =0, 
\end{equation}
To write the above equation in matrix form we define
the matrix $(\mathbb{M_\mu})_{ac}= A_\mu ^b f_{bac}$. Using this defining 
 the killing spinor equation reduces to 
\begin{equation}\label{matrix-eq}
\pd_\mu \epsilon^c + (\mathbb{M_\mu})_{a}^c \epsilon^ a=0. 
\end{equation}
Our task now is to solve (\ref{matrix-eq}). In order to do this
 we make the following ansatz for the solution.
\begin{equation}\label{ansatz-1}
\epsilon = {\cal R}(\rho) e^{\xi x_-} f(x_+), 
\end{equation}
where ${\cal R}(\rho)$ is a square matrix  which is given  by 
\begin{equation}\label{r-dep}
\cal{R}(\rho) = \begin{pmatrix}
e^{-\rho /2} &0 &0 &0 &0 &0  \\
0 &e^{\rho /2} &0 &0 &0 &0  \\
0 &0 &e^{-3\rho /2} &0 &0 &0 \\
0 &0 &0  &e^{-\rho /2} &0 &0 \\
0 &0 &0 &0  &e^{\rho /2} &0 \\
0 &0 &0 &0 &0  &e^{3\rho /2} 
\end{pmatrix}. 
\end{equation}
This ansatz solves the $\rho$ dependence because the matrix $\M_\rho$ has the form\\ Diag$( 1/2,-1/2,3/2,1/2,-1/2,-3/2 )$. 

We now show that connections of the type ({\ref{type-con})  obey the following property :
\begin{equation} \label{claim1}
{\cal R}^{-1} (\rho)  \mathbb{M_\mu} {\R (\rho)} \ \text{is independent of }\rho 
\end{equation}
This can be seen by considering the definitions of 
$\mathbb{M_\mu} $ and ${\cal R}(\rho)$.  Substituting their definitions we obtain 
the following
\begin{align}
\R^{-1}_{ea}  (\M_\mu)_{ac} \R_{cd} &=( e^{-(a)\rho} \delta_{ea}) \ (f_{bac}A_\mu ^b) \ ( e^{(c)\rho} \delta_{cd}),  \nn \\
&= ( e^{-(a)\rho} \delta_{ea}) \ (f_{bac}a_\mu ^b e^{(b)\rho }) \ ( e^{(c)\rho} \delta_{cd}),  \nn \\
&= e^{-(a+b-c)\rho}  \delta_{ea}\delta_{cd}f_{bac}a_\mu ^b \label{prf}. 
\end{align}
Note that the exponential factors  are  negative 
 weights of the corresponding generators.  Since  $[T_b,G_a]\sim G_{a+b}$, 
  $f_{bac}$ is non-zero only when $a+b=c$. Thus the 
   the $\rho$  dependence drops off from (\ref{prf}). 
   This allows us to conclude   that the connections obey the property 
   given in (\ref{claim1}).  Finally we obtain  
\begin{equation}
\R^{-1}_{ea}  (\M_\mu)_{ac} \R_{cd} = f_{bed}a_\mu ^b. 
\end{equation}
Now $f(x_+)$ in (\ref{ansatz-1}) is a column vector which solves the $x_+$ dependence of the Killing spinor. Using the property (\ref{claim1}) in the + component of the Killing spinor equation (\ref{matrix-eq}) we 
obtain
\begin{equation}
\pd _+ f(x_+) + [\R^{-1}(\M _+)\R] f(x_+) =0. 
\end{equation}
Now let  $\lambda_i$ be the eigenvalues of the constant matrix $\R^{-1}(\M _+)\R$,
 then the solution for the above equation is
\begin{equation}
f(x_+)= \sum _i c_i e^{-\lambda_i x_+} \mathsf{z}_{i}, 
\end{equation}
where $\mathsf{z}_i$ is the eigenvector of $\R^{-1}(\M _+)\R$  corresponding to the eigenvalue $\lambda_i$. 
Finally the $x_-$ dependence of the Killing spinor is captured by the simple factor $e^{\xi x_-}$. This is due to the fact that $ (\M _-)_{cd}=-\xi \delta _{cd}  $. 

\subsection{Conical defects}
\subsubsection*{Metric and gauge connections  }

We shall generalize the solution of \cite{Castro:2011iw} to include the spin-1 gauge field  
corresponding to the generators $J$ and the additional spin-2 field corresponding to  the generators 
$V$. We start with the 1-forms, written in terms of the decoupled generators, $T^+$ and $T^-$ as 
defined  in (\ref{decouple-1})
\begin{align}\label{gauge-cone}
A=& (e^{-\rho} \delta_{-1}\tp _{-1} + e^{\rho} \delta_{1}\tp _{1} + e^{-\rho} \beta_{-1}\tm _{-1} + e^{\rho} \beta_{1}\tm _{1} + e^{-\rho} \eta_{-1}W_{-1} + e^{\rho} \eta_{1}W_{1} + \xi J ) dx^+  \nn \\
& \qquad  -\xi J dx^- + (\tm_0 +\tp_0) d\rho,  \\
\bar A=& - (e^{-\rho} \bar \delta_{-1}\tp _{1} + e^{\rho} \bar\delta_{1}\tp _{-1} + e^{-\rho}\bar \beta_{-1}\tm _{1} + e^{\rho} \bar\beta_{1}\tm _{-1} + e^{-\rho} \bar\eta_{-1}W_{1} + e^{\rho} \bar\eta_{1}W_{-1} - \xi J ) dx^-  \nn \\
& \qquad  -\xi J dx^+ - (\tm_0 +\tp_0) d\rho.  
\end{align}
Note that here we have chosen the same notations to 
label the generators in the second copy of $sl(3|2)$ .
Since, $A$ and $\bar A$ are linear combinations of the tetrad $(e)$ and the vielbein $(\omega)$ \cite{Henneaux:2010xg,Campoleoni:2010zq}, we can extract them from the above. The non-zero components of the tetrad turn out to be
\begin{align}
e_\rho & =L_0,  \nn  \\
e_+ & = \tfrac{1}{2} (e^{-\rho} \delta_{-1}\tp_{-1} + e^{\rho} \delta_{1}\tp_{1} + e^{-\rho} \beta_{-1}\tm_{-1} + e^{\rho} \beta_{1}\tm_{1} + e^{-\rho} \eta_{-1}W_{-1} + e^{\rho} \eta_{1}W_{1}  ), \nn  \\
e_- & =  \tfrac{1}{2}  (e^{-\rho} \bar \delta_{-1}\tp_{1} + e^{\rho} \bar\delta_{1}\tp_{-1} + e^{-\rho}\bar \beta_{-1}\tm_{1} + e^{\rho} \bar\beta_{1}\tm_{-1} + e^{-\rho} \bar\eta_{-1}W_{1} + e^{\rho} \bar\eta_{1}W_{-1}  ) .  
\end{align}
The metric is given by the following formula.
\begin{equation}
g_{\mu\nu} = \frac{1}{\epsilon_{(3|2)}} \text{str} (e_\mu e_\nu), 
\end{equation}
where $\epsilon_{(3|2)} = {\rm str} ( L_0^2) = {\rm str} ( \tp_0 + \tm_0)^2$.
Evaluating it explicitly we obtain  $\epsilon_{(3|2)} = 3/4$. By choosing this 
normalization we have chosen the gravitational $sl(2)$ to be the those
corresponding to the generators $L_\pm, L_0$. 
The commutation relations in (\ref{bos}) and (\ref{bosf}) show that it is under
these generators that all fields have well defined weights.  From the 
super ${\cal W}_3$ conformal field theory point of view these are the modes which
are part of the stress tensor of the theory. 
One then obtains
\begin{align}
g_{\rho\rho} &=1\nn,  \\
g_{++} &= -  \tfrac{2}{3}(  \beta_1\beta_{-1} -\tfrac{9}{16} \eta_1\eta_{-1} -\tfrac{1}{4} \delta_1 \delta_{-1}) ,     \\
g_{--} &= -  \tfrac{2}{3} (  \bar\beta_1 \bar\beta_{-1}-\tfrac{9}{16}\bar\eta_1 \bar\eta_{-1}  -\tfrac{1}{4} \bar\delta_1\bar \delta_{-1} ). \nn
\end{align}
We now demand $g_{++}=g_{--}$.  This results in the  following equations
\begin{align}
\bar{\delta}_{\pm 1} = \zeta^{\pm 1} \delta_{\pm 1},  \qquad \bar{\beta}_{\pm 1} = \zeta^{\pm 1} \beta_{\pm 1},  \qquad 
 \bar{\eta}_{\pm 1} = \zeta^{\pm 1} \eta_{\pm 1} . 
\end{align}
where $\zeta$ is constant.  $g_{++}$ and $g_{--}$ now become
\begin{equation}
g_{\pm\pm} =- \tfrac{2}{3}  (  \beta_1\beta_{-1} - \tfrac{9}{16} \eta_1\eta_{-1} -\tfrac{1}{4} \delta_1 \delta_{-1} ), 
\end{equation}
and $g_{+-}$ has the form
\begin{align}
g_{+-}=&\frac{2}{3}\left( - \frac{1}{\zeta} \left( \beta_{-1}^2 - \tfrac{9}{16} \eta_{-1}^2   -\tfrac{1}{4} \delta_{-1}^2 \right)e^{-2\rho}
 -  \zeta   \left( \beta_{1}^2 - \tfrac{9}{16} \eta_{1}^2   -\tfrac{1}{4} \delta_{1}^2 \right)e^{2\rho}\right). 
\end{align}
The metric then in terms of the $(\rho, t, \phi)$ coordinates is as follows.
\begin{align}
ds^2 =& d\rho^2 \nn \\
&- \tfrac{4}{3}\left( \zeta ( \beta_{1}^2 - \tfrac{9}{16} \eta_{1}^2   -\tfrac{1}{4} \delta_{1}^2)e^{2\rho} +2 (  \beta_1\beta_{-1} - \tfrac{9}{16} \eta_1\eta_{-1} -\tfrac{1}{4} \delta_1 \delta_{-1} ) \right. \nn \\
&\left. \hspace{6cm} +\zeta^{-1} ( \beta_{-1}^2 - \tfrac{9}{16} \eta_{-1}^2   -\tfrac{1}{4} \delta_{-1}^2) e^{-2\rho}   \right) dt^2 , \nn \\
& +\tfrac{4}{3}\left( \zeta ( \beta_{1}^2 - \tfrac{9}{16} \eta_{1}^2   -\tfrac{1}{4} \delta_{1}^2)e^{2\rho} -2 (  \beta_1\beta_{-1} - \tfrac{9}{16} \eta_1\eta_{-1} -\tfrac{1}{4} \delta_1 \delta_{-1} ) \right. \nn \\
&\left. \hspace{6cm} +\zeta^{-1} ( \beta_{-1}^2 - \tfrac{9}{16} \eta_{-1}^2   -\tfrac{1}{4} \delta_{-1}^2) e^{-2\rho}   \right)   d\phi^2.  \nn
\end{align}
We now need to impose the fact that $g_{tt}$ and $g_{\phi\phi}$ need to have a perfect square form. 
The results in the following equation
\begin{align}
( \beta_1^2 -\tfrac{9}{16}\eta_1^2 -\tfrac{1}{2} \delta_1^2 ) &(  \beta_{-1}^2 - \tfrac{9}{16}\eta_{-1}^2-\tfrac{1}{2} \delta_{-1}^2)  = ( \beta_1\beta_{-1} -\tfrac{9}{16}\eta_1\eta_{-1}-\tfrac{1}{4}\delta_1\delta_{-1}	 )^2
\end{align}
This imposes the conditions 
\begin{align}
\delta_{-1} = \alpha \delta_1 ,\qquad \beta_{-1}= \alpha\beta_1 , \qquad \eta_{-1}=\alpha\eta_{1}.
\end{align}
Defining $\delta=\delta_1,\ \beta=\beta_1$ and$ \ \eta=\eta_1$, the final form the metric with these conditions is
\begin{align}\label{conical-metric}
ds^2 =& \, d\rho^2 - 	\tfrac{4}{3}( \beta^2 -(\tfrac{3}{4}\eta)^2 -(\tfrac{1}{2} \delta)^2 )  \left[ \left( \sqrt{\zeta}e^\rho + \frac{\alpha}{\sqrt{\zeta}} e^{-\rho} \right)^2 dt^2 - \left( \sqrt{\zeta}e^\rho - \frac{\alpha}{\sqrt{\zeta}} e^{-\rho} \right)^2 d\phi ^2 \right]. 
\end{align}
By redefining $\rho$ as $\rho \rightarrow \rho - \frac{1}{2}\log\left( \frac{\zeta}{\alpha} \right)$ we can write (\ref{conical-metric}) as
\begin{align}\label{conical-metric2}
ds^2 =& \, d\rho^2 - \tfrac{16\alpha}{3}	( \beta^2 -(\tfrac{3}{4}\eta)^2-(\tfrac{1}{2} \delta)^2   ) \left[(\sinh ^2 \rho)  dt^2 - (\cosh ^2 \rho ) d\phi ^2 \right]. 
\end{align}
From examining this metric it is easy to see that it is only for special values of the 
parameters $\alpha, \beta, \eta, \delta$ the metric reduces to global $AdS_3$. 
For generic values the solution is metrically  singular. 
The special values at which these solutions reduce to solutions studied earlier 
in the literature will be discussed subsequently.

\subsubsection*{Killing spinors for the higher spin conical defect}

The equation for the covariantly constant spinor  is  given by 
\begin{equation}\label{cov-hscdx}
\cD_\mu \lambda \equiv \pd_\mu  \lambda + [A_\mu , \lambda] =0, 
\end{equation}
where $\lambda$ is given by
\begin{equation}
\lambda \equiv \sum_{r=-1/2}^{1/2}  \e ^{r}G^+_{r} + \sum_{r=-1/2}^{1/2}  \be ^{r}G^-_{r} +\sum_{r=-3/2}^{3/2}  \l ^{r}U^+_{r} + \sum_{r=-3/2}^{3/2}  \tl ^{r}U^-_{r} . 
\end{equation}
From the analysis of the previous section the gauge connection for the 
higher conical defect is given by  
\begin{align} \label{fincon}
A= &\  (\alpha \delta e^{-\rho} \tp_{-1} + \delta  e^\rho \tp_1 + \alpha\beta  e^{-\rho} \tm_{1} + \beta  e^{\rho} \tm_{-1} + \alpha\eta  e^{-\rho} W_{-1} + \eta e^\rho W_1 + \xi J) dx^+ \nn \\
& \qquad   -\xi J_0 dx^- +L_0 d\rho . 
\end{align}
where $L_0 = \tp_0 + \tm_0$. 
We will study  the supersymmetry of only  one copy of
 the $sl(3|2)_L \times sl(3|2)_R$
 Chern-Simons theory. A similar analysis can be repeated for the second copy.

Extracting  out  the components of the connection given in (\ref{fincon}) 
 as in (\ref{type-con}) along with the exponential $\rho$ dependence  we obtain
\begin{eqnarray}
& & j _+ = \xi , \quad j_-  = -\xi   \\
& & s_+^{-1} = \alpha\delta  e^{-\rho}, \quad s_+^{1} = \delta  e^{\rho}, \quad l_\rho^0=1,   \nn \\
& & t_+^{-1} =\alpha\beta e^{-\rho} , \quad t_+^1=\beta   e^\rho,  \nn \\
& &w_+^{-1} = \alpha\eta  e^{-\rho} , \quad w_+^1 = \eta e^\rho,  \nn 
\end{eqnarray}
where, $t$ and $s$ are the components corresponding to the generators $\tp$ and $\tm$ respectively. 
The  equation (\ref{cov-hscdx}) for the components $G^+_r$ and $U^+_r$ 
is given by 
\begin{scriptsize}
\arraycolsep=0pt
\medmuskip = 0mu
\begin{align}
&\pd _\mu  \begin{pmatrix}
\e^{-1/2} \\
\e^{1/2} \\
\l^{-3/2} \\
\l^{-1/2} \\
\l^{1/2} \\
\l^{3/2}
\end{pmatrix}\nn \\
&+\frac{1}{4} \begin{pmatrix}
4j_0 +2l_0 &\tfrac{16}{3}s_{-1}+\frac{4}{3}t_{-1} &4i(s_1-t_1)+3w_1 &0 &\frac{4i}{3}(s_{-1}-t_{-1})+3w_{-1} &0 \\
\frac{16}{3}s_1-\frac{4}{3}t_1 &4j_0 -2l_0 &0 &\frac{4i}{3}(s_1-t_1)-3w_1 &0 &4i(s_{-1}-t_{-1})-3w_1 \\
\frac{8i}{3} (s_{-1}-t_{-1})+2w_{-1} &0 &4j_0 +6l_0 &-\frac{8}{3}s_{-1}-\frac{4}{3}t_{-1}+2iw_{-1} &0 &0 \\
0 &\frac{8i}{3}(s_{-1}-t_{-1})-6w_{-1} &8s_1 +4t_1-6iw_1 &4j_0+2l_0 &-\frac{16}{3}s_{-1}-\frac{8}{3}t_{-1} &0 \\
\frac{8i}{3}(s_1-t_1)+6w_1 & 0 &0 &\frac{16}{3}s_1 +\frac{8}{3}t_1 &4j_0-2l_0 &-8s_{-1}-4t_{-1}-6iw_{-1} \\
0 &\frac{8i}{3}(s_1-t_1)-2w_1 &0 &0 &\frac{8}{3}s_1+\frac{4}{3}t_1 +2iw_1 &4j_0 -6l_0
\end{pmatrix}  _{\mu} \nn \\
&\qquad\qquad\qquad \times   \begin{pmatrix}
\e^{-1/2} \\
\e^{1/2} \\
\l^{-3/2} \\
\l^{-1/2} \\
\l^{1/2} \\
\l^{3/2}
\end{pmatrix} =0. 
\end{align}
\end{scriptsize}
The $x^+$ dependence of the column spinor above is determined by the eigenvalues of the $\R ^{-1} (\rho) \M_+ \R(\rho)$ matrix. The solutions are of the form 
\begin{align}
\begin{pmatrix}
\e^{-1/2} \\
\e^{1/2} \\
\l ^{-3/2} \\
\l ^{-1/2} \\
\l ^{1/2}\\
\l ^{3/2}
\end{pmatrix}  \nn \\ &= \mathcal{R}(\rho)\, e^{\xi(x_- - x_+)} (d_1 e^{{i\sqrt{\alpha}\delta  } x_+} \mathsf{z} + d_2 e^{{-i\sqrt{\alpha}\delta } x_+} \mathsf{z}_2 \nn \\
 &\qquad +d_3 e^{{i\sqrt{\alpha}\left(\delta  +2\left( \beta^2-(\tfrac{3}{4}\eta)^2  \right)^{1/2}\right)} x_+} \mathsf{z}_3 +  d_4 e^{-{i\sqrt{\alpha}\left(\delta   +2\left( \beta^2-(\tfrac{3}{4}\eta)^2  \right)^{1/2}\right)} x_+} \mathsf{z}_4  \nn \\
  &\qquad +d_5 e^{{i\sqrt{\alpha}\left(\delta  -2\left( \beta^2-(\tfrac{3}{4}\eta)^2  \right)^{1/2}\right)} x_+} \mathsf{z}_5 +  d_6 e^{-{i\sqrt{\alpha}\left(\delta  -2\left( \beta^2-(\tfrac{3}{4}\eta)^2  \right)^{1/2}\right)} x_+} \mathsf{z}_6 ) . 
\end{align}
The matrix $\mathcal{R}$ has the $\rho$ dependence as in (\ref{r-dep}).

Now re-expressing $x_+$ and $x_-$ in terms of the co-ordinates $t$ and $\phi$ 
allows us to obtain the condition under which any of  the above  Killing spinor is periodic. 
 The possibilities are the following: 
\begin{align}\label{spinor-cond-1}
2\xi &= \pm i\sqrt{\alpha} \delta +in,  \\
2\xi &= \pm i\sqrt{\alpha} \left(   \delta \pm 2 (\beta ^2 - (\tfrac{3}{4}\eta)^2 )^{1/2}  \right) +in. 
 \label{spinor-cond-2}
\end{align}
where $n$ is any integer. 

We have also  examined the Killing spinor equation for the $G^-$ and $U^-$ components. 
On repeating  the same analysis we have seen   that the component $u(1)$ gauge field,
$\xi$ has to be complex in order to impose proper periodicity requirements. 
Since this is not allowed we conclude that there are no  Killing spinors corresponding to conjugates of 
the $G^-, U^-$ charges. 

To relate to  known solutions,  we will obtain  the special values of the parameter space
at which  the 
higher spin conical defect reduces to  solutions studied earlier in the literature. 

\subsubsection*{Supersymmetry of conical defects in $sl(2)$ }

Embedding the conical defect solution only in 
 the $sl(2)\oplus u(1)$ sub-algebra we have the following  gauge connections
\begin{align}\label{simple-cone}
A &= \left( e^\rho \tp_1 + \frac{\gamma}{4} e^{-\rho} \tp_{-1} \right)dx^+ +\tp_0 d\rho + 2\xi Jd\phi,   \\
\bar A &= -\left( e^\rho \tp_{-1} + \frac{\gamma}{4} e^{-\rho} \tp_{1} \right)dx^+ - \tp_0 d\rho + 2\xi J d\phi.  \nn
\end{align}
Note that this gauge connection
is a special case of the higher spin conical defect with $\alpha=\gamma /4$,  $ \delta =1 $ and 
$\beta =\eta=0$.

One can perform a gauge transformation $A\rightarrow U^-(A+d)U$ with $U=e^{\rho \tm_0}$ on the connection (\ref{simple-cone}). The new connections are then of the form
\begin{align}\label{simple-cone2}
A &= \left( e^\rho \tp_1 + \frac{\gamma}{4} e^{-\rho} \tp_{-1} \right)dx^+ +(\tp_0+ \tm_0) d\rho + 2\xi J d\phi,   \\
\bar A &= -\left( e^\rho \tp_{-1} + \frac{\gamma}{4} e^{-\rho} \tp_{1} \right)dx^+ - 
(\tp_0+ \tm_0) d\rho + 2\xi J d\phi.  \nn
\end{align}
where, for $\bar A$ we have used the transformation by $U=e^{-\rho \tm_0}$.
Now the gauge connections are of the general form given in  (\ref{type-con}). 

The equation for the covariantly constant spinor is 
\begin{equation}\label{cov-hscd}
\cD_\mu \lambda \equiv \pd_\mu  \lambda + [A_\mu , \lambda] =0, 
\end{equation}
where $\lambda$ is given by
\begin{equation}
\lambda \equiv \sum_{r=-1/2}^{1/2}  \e ^{r}G^+_{r} + \sum_{r=-1/2}^{1/2}  \be ^{r}G^-_{r} +\sum_{r=-3/2}^{3/2}  \l ^{r}U^+_{r} + \sum_{r=-3/2}^{3/2}  \tl ^{r}U^-_{r} . 
\end{equation}
The  analysis for the Killing spinor performed for   the case of the higher spin conical defect 
can be repeated. The solutions of the components  of the  generators 
$G^\pm, U^\pm$  are of the form 
\begin{align}&
\begin{pmatrix}
\e^{-1/2} \\
\e^{1/2} \\
\l ^{-3/2} \\
\l ^{-1/2} \\
\l ^{1/2}\\
\l ^{3/2}
\end{pmatrix} _{\pm} \nn \\& = \mathcal{R}(\rho)\, e^{\xi(x_- - x_+)} \left( e^{i\frac{\sqrt{\gamma}}{2} x_+}(d_1 \mathsf{z}_1 +d_2 \mathsf{z}_2+d_3 \mathsf{z}_3   )+  e^{{-i\frac{\sqrt{\gamma}}{2} } x_+}(d_4 \mathsf{z}_4 + d_5 \mathsf{z}_5 +d_6 \mathsf{z}_6)\right)_{\pm}
\end{align}
$\mathsf{z}_{i}$ are the eigenvectors of the $6\times6$ matrices which appear in the Killing spinor equation. The $\rho$ dependence in contained in the matrix $\mathcal{R}(\rho)$ given in (\ref{r-dep}).

Imposing   periodicity  on the  Killing spinor, we obtain 
\begin{equation} \label{eigc}
\xi = \pm i \frac{\sqrt{\gamma}}{4}+in. 
\end{equation}
Note that this condition coincides with the condition found for Killing spinors in  
\cite{Izquierdo:1994jz}. 
Since there is a  pair of eigenvalues with degeneracy $3$,  we will in general  have 
$3$ Killing spinors which will satisfy the periodicity condition.

\subsubsection*{Supersymmetry of Anti-deSitter space in $sl(2)$ }

For the case of $AdS_3$ one can perform the same analysis with $\gamma=1$.
As expected,   it  can be seen that one  does not require the $u(1)$ gauge field and 
one  obtains anti-periodic Killing spinors.  The solution for the Killing spinors for this case is
\begin{align}
\begin{pmatrix}
\e^{-1/2} \\
\e^{1/2} \\
\l ^{-3/2} \\
\l ^{-1/2} \\
\l ^{1/2}\\
\l ^{3/2}
\end{pmatrix} _{\pm}  = \mathcal{R}(\rho)\,  \left( e^{i\frac{x_+}{2} }   ( d_1 \mathsf{z}_1 + d_2 \mathsf{z}_2 +d_3 \mathsf{z}_3)+  e^{-i\frac{x_+}{2} }(d_4 \mathsf{z}_4 + d_5 \mathsf{z}_5 +d_6 \mathsf{z}_6) \right) _{\pm}. 
\end{align}
The $\rho$ dependence of the Killing spinor remains the same as the one for the conical defect. 
This $AdS_3$  in $sl(2)$ admits $6$ anti-periodic Killing spinors.

\subsubsection*{Supersymmetry of conical defects  in  the gravitational $sl(2)$}

We now write down the 
metric for the conical defect embedded in the gravitational $sl(2)$  generated 
by the $L_m$ generators. In  Fefferman-Graham coordinates this metric is given by 
\begin{equation}
ds^2 = d\rho^2 - \left( e^\rho + \frac{\gamma}{4} e^{-\rho} \right) ^2 dt^2 +  \left( e^\rho - \frac{\gamma}{4} e^{-\rho} \right) ^2 d\phi^2 . 
\end{equation}
This can be  equivalently written in terms of the gauge connections 
\begin{align}\label{conL}
A &= \left( e^\rho L_1 + \frac{\gamma}{4} e^{-\rho} L_{-1} \right)dx^+ +L_0 d\rho + 2\xi J_0 d\phi,   \\
\bar A &= -\left( e^\rho L_{-1} + \frac{\gamma}{4} e^{-\rho} L_{1} \right)dx^+ - L_0 d\rho + 2\xi J_0 d\phi. 
\end{align}
Note that this connection is a special case of (\ref{fincon}) with $\beta = \delta ,\eta =0, \zeta=1$
and $\alpha = \frac{\gamma}{4}$. 
 These connections and the metric reduce to that of global AdS by setting $\gamma=1$ and $\xi=0$.  
The  non-zero components of the connection are 
\begin{align}
l_+^1&=e^\rho\ , \qquad l_+^{-1}=\frac{\gamma}{4} e^{-\rho} \ , \qquad l_\rho^0 = 1 \ ,
&j_+^0 = \xi \ , \qquad j^0_- = -\xi \ . 
\end{align}

The equation for the covariantly conserved spinor is  given by 
\begin{equation}
\cD_\mu \lambda \equiv \pd_\mu  \lambda + [A_\mu , \lambda] =0,  
\end{equation}
where $\lambda$ is given by
\begin{equation}
\lambda \equiv \sum_{r=-1/2}^{1/2}  \e ^{r}G^+_{r} + \sum_{r=-1/2}^{1/2}  \be ^{r}G^-_{r} +\sum_{r=-3/2}^{3/2}  \l ^{r}U^+_{r} + \sum_{r=-3/2}^{3/2}  \tl ^{r}U^-_{r} . 
\end{equation}
For the  connection given in  (\ref{conL})  the Killing spinor equations for the $G^\pm _r$ and $U^\pm _r$ decouple. The equations for the $G^+_{\pm 1/2}$ components  in matrix form  is given by 
\begin{equation}\label{m01}
\pd _\mu \begin{pmatrix}
 \e ^{-1/2} \\
  \e ^{1/2} 
\end{pmatrix} + \begin{pmatrix}
\frac{1}{2} (2j^0 +l^0)_\mu  & - l^{-1}_\mu \\
 l^{1}_\mu  & \frac{1}{2} (2j^0 -l^0)_\mu
\end{pmatrix}  \begin{pmatrix}
 \e ^{-1/2} \\
  \e ^{1/2} 
\end{pmatrix} =0 . 
\end{equation}
The solutions are given by 
\begin{align} \label{e11}
\begin{pmatrix}
\e ^{-1/2} \\
\e ^{1/2}
\end{pmatrix} &= \mathcal{R}_1(\rho) \, e^{\xi( x_- - x_+)}(c_1 e^{i\tfrac{\sqrt{\gamma}}{2} x_+} \mathsf{y}_1 + c_2 e^{-i\tfrac{\sqrt{\gamma}}{2} x_+} \mathsf{y}_2) \nn \\
&=  \mathcal{R}_1(\rho) \, e^{-2\xi\phi}(c_1 e^{i\tfrac{\sqrt{\gamma}}{2}(t+\phi)} \mathsf{y}_1 + c_2 e^{-i\tfrac{\sqrt{\gamma}}{2}(t+\phi)} \mathsf{y}_2),   
\end{align} 
where, $\mathsf{y}_{1,2}$ are the eigenvectors of the   matrix  $\R^{-1}_1\M_+\R_1$ . Here
$\M_\mu$ is  the matrix which appears in  the equation (\ref{m01}) and  $ \mathcal{R}_1(\rho) $ is a diagonal matrix with the following  $\rho$ dependence
\begin{equation}
\mathcal{R}_1(\rho) =  \begin{pmatrix}
e^{-\rho /2}  & 0 \\
0 & e^{\rho /2}
\end{pmatrix}. 
\end{equation}
The  equations for the $U^+_r$ generators are 
\begin{align}
\pd _\mu \begin{pmatrix}
\l ^{-3/2} \\
\l ^{-1/2} \\
\l ^{1/2}\\
\l ^{3/2}
\end{pmatrix} + \begin{pmatrix}
\tfrac{1}{2}(2j^0 +3 l^0)_\mu   &- l^{-1}_\mu &0 &0 \\
3l^1_\mu &\tfrac{1}{2}(2j^0 + l^0)_\mu &-2l^{-1}_\mu &0 \\
0  &2l^1_\mu  &\tfrac{1}{2}(2j^0 - l^0)_\mu &-3l^{-1}_\mu \\
0 &0 &l^1_\mu &\tfrac{1}{2}(2j^0 -3 l^0)_\mu
\end{pmatrix}  \begin{pmatrix}
\l ^{-3/2} \\
\l ^{-1/2} \\
\l ^{1/2}\\
\l ^{3/2}
\end{pmatrix}  &= 0. 
\end{align} 
The solutions are  given by  
\begin{align}\label{l11}
 \begin{pmatrix}
\l ^{-3/2} \\
\l ^{-1/2} \\
\l ^{1/2}\\
\l ^{3/2}
\end{pmatrix}& \nn \\
&= \mathcal{R}_2(\rho)\, e^{\xi (x_- -x_+)}(d_1 e^{i\tfrac{\sqrt{\gamma}}{2} x_+} \mathsf{z}_1 + d_2 e^{-i\tfrac{\sqrt{\gamma}}{2} x_+} \mathsf{z}_2  \nn \\
&\hspace{4cm}  + d_3 e^{i\tfrac{3\sqrt{\gamma}}{2} x_+} \mathsf{z}_3 + d_4 e^{-i\tfrac{3\sqrt{\gamma}}{2} x_+} \mathsf{z}_4 ),  \nn \\
&=  \mathcal{R}_2(\rho) \, e^{-2\xi\phi}(d_1 e^{i\tfrac{\sqrt{\gamma}}{2}(t+\phi)} \mathsf{z}_1 + d_2 e^{-i\tfrac{\sqrt{\gamma}}{2}(t+\phi)} \mathsf{z}_2 \nn \\
&\hspace{4cm} + d_3 e^{i\tfrac{3\sqrt{\gamma}}{2}(t+\phi)} \mathsf{z}_3 + d_4 e^{-i\tfrac{3\sqrt{\gamma}}{2}(t+\phi)} \mathsf{z}_4   ).  
\end{align}
The matrix $\mathcal{R}_2$ has the $\rho$ dependence
\begin{equation}
\mathcal{R}_2 (\rho)= \begin{pmatrix}
e^{-3\rho /2} &0 &0 &0 \\
0 &e^{-\rho /2} &0 &0  \\
0 &0 &e^{\rho /2} &0 \\
0 &0 &0 &e^{3\rho /2}  
\end{pmatrix}. 
\end{equation}
We thus get 6 independent Killing spinors. The conditions which we obtain
 on demanding periodicity of the spinor is 
\begin{align}
\xi= \pm i \frac{\gamma}{4}+in, \qquad \text{or} \qquad \xi= \pm 3i \frac{\gamma}{4}+in.
\end{align}
Thus, on embedding this conical defect in the $sl(2)$ corresponding to $L_0, L_\pm$ we see
 that there are $4$ eigenvalues  out of which 
there are $2$ doubly degenerate ones.  The doubly degenerate ones obey the condition
$\xi= \pm i \frac{\gamma}{4}+in$. 
 These match with 
that given in (\ref{eigc}) which also agrees with \cite{Izquierdo:1994jz}.

The Killing spinor equations for the $G^-_r$ and $U^-_r$ components of  also form a set of 6 coupled equations. These equations are the same as the above with the replacement $j_0 \rightarrow -j_0$ or $\xi \rightarrow -( -\xi)$. Thus, they admit same solutions as given in (\ref{e11}) and (\ref{l11})
 with  different arbitrary constants 

\subsubsection*{Supersymmetry of anti-de Sitter space  in  the gravitational $sl(2)$}

Let us now turn to the case of  global  {$AdS_3$ }  embedded in  the gravitational $sl(2)$. 
This is a special case of the conical 
spaces embedded in the gravitational 
$sl(2)$  with $\gamma =1, \, \xi=0 $. The metric in terms of the Fefferman-Graham coordinates is
\begin{equation}
ds^2 = d\rho^2 - \left( e^\rho + \frac{1}{4} e^{-\rho} \right) ^2 dt^2 +  \left( e^\rho - \frac{1}{4} e^{-\rho} \right) ^2 d\phi^2 . 
\end{equation}
This can be  equivalently written in terms of the gauge connections 
\begin{align}
A &= \left( e^\rho L_1 + \frac{1}{4} e^{-\rho} L_{-1} \right)dx^+ +L_0 d\rho ,  \\ \nonumber
\bar A &= -\left( e^\rho L_{-1} + \frac{1}{4} e^{-\rho} L_{1} \right)dx^+ - L_0 d\rho. 
\end{align}
The solutions for the Killing spinors are  given by 
\begin{align}
\begin{pmatrix}
\e ^{-1/2} \\
\e ^{1/2}
\end{pmatrix} &= \mathcal{R}_1(\rho) \, (c_1 e^{\tfrac{{i}}{2} x_+} \mathsf{y}_1 + c_2 e^{-\tfrac{{i}}{2} x_+} \mathsf{y}_2),  \nn \\
&=  \mathcal{R}_1(\rho) \, (c_1 e^{\tfrac{{i}}{2}(t+\phi)} \mathsf{y}_1 + c_2 e^{-\tfrac{{i}}{2}(t+\phi)} \mathsf{y}_2),   \label{ads-sol-1}
\end{align} 
and
\begin{align}
\begin{pmatrix}
\l ^{-3/2} \\
\l ^{-1/2} \\
\l ^{1/2}\\
\l ^{3/2}
\end{pmatrix} &= \mathcal{R}_2(\rho)\, (d_1 e^{\tfrac{i}{2} x_+} \mathsf{z}_1 + d_2 e^{-\tfrac{i}{2} x_+} \mathsf{z}_2 + d_3 e^{\tfrac{3i}{2} x_+} \mathsf{z}_3 + d_4 e^{-\tfrac{3i}{2} x_+} \mathsf{z}_4 ),  \nn \\
&=   \mathcal{R}_2(\rho)\, (d_1 e^{\tfrac{i}{2}  (t+\phi)} \mathsf{z}_1 + d_2 e^{-\tfrac{i}{2} (t+\phi)} \mathsf{z}_2 + d_3 e^{\tfrac{3i}{2}  (t+\phi)} \mathsf{z}_3 + d_4 e^{-\tfrac{3i}{2}  (t+\phi)} \mathsf{z}_4 ) . 
\end{align}
We obtain $6$ independent Killing spinors which are anti-periodic corresponding to the 
 $G^+_r$ and $U^+_r$ generators.  Similarly performing the same analysis  it 
is easy to see that one obtains $6$ independent anti-periodic  Killing spinors 
corresponding to the  $G^-_r$ and $U^-_r$ generators.

The holonomy of  global $AdS_3$ around the angular direction
$\phi$  embedded in the gravitational $sl(2)$ 
 can be shown  to be trivial and therefore the solution is smooth. 
Thus this background 
corresponds to  the supersymmetric vacuum in the Neveu-Schwarz sector of the dual CFT.

\subsection{The BTZ black hole}
\subsubsection*{The BTZ black hole in $sl(2)$ }
We now examine the supersymmetry of the connection corresponding to that
of the BTZ black hole embedded in the $sl(2)$ part of bosonic algebra $sl(3)\oplus sl(2) \oplus u(1)$. 
The  connections are given by 
\begin{align}
A &= \left(  e^\rho \tp _1 -\frac{2\pi}{k}\cL e^{-\rho}  \tp _{-1}  \right) dx^+ +\tp _0 d\rho,  \\ \nonumber
\bar{A}  &= -\left(  e^\rho \tp _{-1} -\frac{2\pi}{k}\bar{\cL}e^{-\rho} \tp  _{1}  \right) dx^- -\tp _0 d\rho, 
\end{align}
where
\begin{equation}
\cL = \frac{M-\hat J}{4\pi}  \quad , \quad \bar \cL = \frac{M+\hat J}{4\pi}
\end{equation}
We shall make a gauge transformation $A\rightarrow U^-( A+d) U$ to the above connections with $U=e^{\rho \tm_0}$ for $A$ and $U=e^{-\rho \tm_0}$ for $\bar A$. This gives
\begin{align}
A &= \left(  e^\rho \tp _1 -\frac{2\pi}{k}\cL e^{-\rho}  \tp _{-1}  \right) dx^+  +
( \tp_0 + \tm_0)d\rho,  \\ \nonumber
\bar{A}  &= -\left(  e^\rho \tp _{-1} -\frac{2\pi}{k}\bar{\cL}e^{-\rho} \tp  _{1}  \right) dx^- -( \tp_0 + \tm_0) d\rho. 
\end{align}
Now the connection is of the  general form  given by (\ref{type-con}). 

For the extremal case ($M=\hat J$) we have, $\cL=0$ and therefore the connection $A$ reduces to 
\begin{equation}
A = e^\rho \tp_1  dx^+ + ( \tp_0 + \tm_0) d\rho. 
\end{equation}
The equation for the Killing spinor is  given by 
\begin{equation}\label{cov1}
\cD_\mu \lambda \equiv \pd_\mu  \lambda + [A_\mu , \lambda] =0, 
\end{equation}
where $\lambda$ is expanded as 
\begin{equation}
\lambda \equiv \sum_{r=-1/2}^{1/2}  \e ^{r}G^+_{r} + \sum_{r=-1/2}^{1/2}  \be ^{r}G^-_{r} +\sum_{r=-3/2}^{3/2}  \l ^{r}U^+_{r} + \sum_{r=-3/2}^{3/2}  \tl ^{r}U^-_{r}.  
\end{equation}
Writing the equation given in   (\ref{cov1}) explicitly we obtain the following 
equation 
  for the $\rho$ direction 
\begin{align}
\pd _\rho   \begin{pmatrix}
\e^{-1/2} \\
\e^{1/2} \\
\l^{-3/2} \\
\l^{-1/2} \\
\l^{1/2} \\
\l^{3/2}
\end{pmatrix} + \begin{pmatrix}
\tfrac{1}{2} &0 &0 &0 &0 &0 \\
0 &-\tfrac{1}{2}  &0 &0 &0 &0 \\
0 &0 &\tfrac{3}{2}  &0 &0 &0  \\
0 &0 &0  &\frac{1}{2} &0 &0 \\
0 &0 &0 &0  &-\frac{1}{2}  &0 \\
0 &0 &0 &0  &0  &-\tfrac{3}{2}
\end{pmatrix}  \begin{pmatrix}
\e^{-1/2} \\
\e^{1/2} \\
\l^{-3/2} \\
\l^{-1/2} \\
\l^{1/2} \\
\l^{3/2}
\end{pmatrix} =0.
\end{align}
Similarly the equation for   the $+$ direction is given by 
\begin{align}
\pd _+   \begin{pmatrix}
\e^{-1/2} \\
\e^{1/2} \\
\l^{-3/2} \\
\l^{-1/2} \\
\l^{1/2} \\
\l^{3/2}
\end{pmatrix} +  \left(
\begin{array}{cccccc}
 0 & 0 & -i & 0 & 0 & 0 \\
 -\frac{1}{3} & 0 & 0 & -\frac{i}{3} & 0 & 0 \\
 0 & 0 & 0 & 0 & 0 & 0 \\
 0 & 0 & 1 & 0 & 0 & 0 \\
 -\frac{2 i}{3} & 0 & 0 & \frac{2}{3} & 0 & 0 \\
 0 & -\frac{2 i}{3} & 0 & 0 & \frac{1}{3} & 0
\end{array}
\right) \begin{pmatrix}
\e^{-1/2} \\
\e^{1/2} \\
\l^{-3/2} \\
\l^{-1/2} \\
\l^{1/2} \\
\l^{3/2}
\end{pmatrix} =0. 
\end{align}
The solutions of these equations are  of the form
\begin{equation}
\begin{pmatrix}
\e^{-1/2} \\
\e^{1/2} \\
\l^{-3/2} \\
\l^{-1/2} \\
\l^{1/2} \\
\l^{3/2}
\end{pmatrix} = \begin{pmatrix}
c_3 e^{-\rho /2 -i\pi /2} \\
\frac{c_2}{2}e^{\rho /2 -i\pi /2} \\
0 \\
c_3 e^{-\rho /2} \\
c_2 e^{\rho /2} \\
c_1 e^{3\rho /2}  
\end{pmatrix}. 
\end{equation}
Thus there are $3$ linearly independent Killing spinors corresponding to the supercharges with  positive $\hat J$ charge  for the extremal BTZ embedded in  the $sl(3|2)$ theory.

\subsubsection*{The BTZ black hole in gravitational $sl(2)$ }

The connection of the
BTZ black hole  embedded in the gravitational $sl(2)$  is given by 
\begin{align}
A &= \left(  e^\rho L_1 -\frac{2\pi}{k}\cL e^{-\rho}  L_{-1}  \right) dx^+ +L_0 d\rho,  \\ \nonumber
\bar{A}  &= -\left(  e^\rho L_{-1} -\frac{2\pi}{k}\bar{\cL}e^{-\rho}  L_{1}  \right) dx^- -L_0 d\rho, 
\end{align}
where
\begin{equation}
\cL = \frac{M-\hat J}{4\pi}  \quad , \quad \bar \cL = \frac{M+\hat J}{4\pi}. 
\end{equation}
Substituting $M=\hat J$  for the extremal BTZ the connection reduces to 
\begin{equation}
A = e^\rho L_1  dx^+ +L_0 d\rho. 
\end{equation}

The equation for the covariantly constant  spinor is  given by 
\begin{equation}\label{cov2}
\cD_\mu \lambda \equiv \pd_\mu  \lambda + [A_\mu , \lambda] =0, 
\end{equation}
where $\lambda$ is expanded as 
\begin{equation}
\lambda \equiv \sum_{r=-1/2}^{1/2}  \e ^{r}G^+_{r} + \sum_{r=-1/2}^{1/2}  \be ^{r}G^-_{r} +\sum_{r=-3/2}^{3/2}  \l ^{r}U^+_{r} + \sum_{r=-3/2}^{3/2}  \tl ^{r}U^-_{r} . 
\end{equation}
Writing out the Killing spinor equations 
 for the $G^+_r$ generators we obtain
\begin{equation}
\pd _\mu \begin{pmatrix}
 \e ^{-1/2} \\
  \e ^{1/2} 
\end{pmatrix} + \begin{pmatrix}
\frac{1}{2} (2j^0 +l^0)_\mu  & - l^{-1}_\mu \\
- l^{1}_\mu  & \frac{1}{2} (2j^0 -l^0)_\mu
\end{pmatrix}  \begin{pmatrix}
 \e ^{-1/2} \\
  \e ^{1/2} 
\end{pmatrix} =0 . 
\end{equation}
Similarly the equations  for $U^+$ generators are given by 
\begin{align}
\pd _\mu \begin{pmatrix}
\l ^{-3/2} \\
\l ^{-1/2} \\
\l ^{1/2}\\
\l ^{3/2}
\end{pmatrix} + \begin{pmatrix}
\tfrac{3}{2} l^0 _\mu   &0 &0 &0 \\
3l^1_\mu &\tfrac{3}{2} l^0 _\mu &0 &0 \\
0  &2l^1_\mu  &-\tfrac{3}{2}  l^0 _\mu &0 \\
0 &0 &l^1_\mu &-\tfrac{3}{2}  l^0_\mu
\end{pmatrix}  \begin{pmatrix}
\l ^{-3/2} \\
\l ^{-1/2} \\
\l ^{1/2}\\
\l ^{3/2}
\end{pmatrix}  &= 0. 
\end{align} 
The  equations for $\be$ and $\tl$ are identical to these but 
with the replacements $\e \rightarrow \be$ and 
$\l \rightarrow \tl$. 
The Killing spinor which is periodic in the angular direction is given  by
\begin{align} \label{btz-ans}
\begin{pmatrix}
 \e^{-1/2} \\
  \e^{1/2}
\end{pmatrix}  = \begin{pmatrix}
0 \\
Ce^{\rho/2}
\end{pmatrix}  \quad &, \quad \begin{pmatrix}
 \be^{-1/2} \\
  \be^{1/2}
\end{pmatrix}  = \begin{pmatrix}
0 \\
\tilde  Ce^{\rho/2}
\end{pmatrix} \\
\begin{pmatrix}
  \l^{-3/2} \\
 \l^{-1/2} \\
  \l^{1/2} \\
   \l^{3/2}
\end{pmatrix}  = \begin{pmatrix}
0\\
0\\
0 \\
D e^{3\rho/2}
\end{pmatrix}  \quad &, \quad \begin{pmatrix}
  \tl^{-3/2} \\
 \tl^{-1/2} \\
  \tl^{1/2} \\
   \tl^{3/2}
\end{pmatrix}  = \begin{pmatrix}
0\\
0\\
0 \\
\tilde D e^{3\rho/2}
\end{pmatrix} 
\end{align}
The solution given in  (\ref{btz-ans}) matches with that  obtained by  \cite{Tan:2012xi}. 
Thus the extremal BTZ embedded in the gravitational $sl(2)$ admits $2$ independent Killing
spinors corresponding to the $G^+, U^+$ generators. 

\subsection{Higher spin black holes}

Now we will study the supersymmetry  of  black holes with spin-3 charge   recently constructed 
 in \cite{Gutperle:2011kf}.  The connections are given by 
\begin{align}\label{w3bh-c}
A=& \left( e^\rho  \tm_1 - \frac{2\pi}{k}\cL e^{-\rho}\tm_{-1} + \frac{\pi}{2k\sigma} \cW e^{-2\rho}W_{-2}  \right) dx^+ \nn \\
     & +\mu\left( e^{2\rho} W_{-2} -\frac{4\pi \cL}{k} W_0 + \frac{4\pi^2\cL^2}{k^2}e^{-2\rho}W_{2}+\frac{4\pi \cW}{k} e^{-\rho}\tm_{-1} \right) dx^{-} + 2\xi J  d\phi +L_0 d\rho,  \\
\bar A=& - \left( e^\rho  \tm_1 - \frac{2\pi}{k} \bar\cL e^{-\rho}\tm_{-1} + \frac{\pi}{2k\sigma} \bar\cW e^{-2\rho}W_{-2}  \right) dx^+  \nn \\
     & -\bar\mu\left( e^{2\rho} W_{-2} -\frac{4\pi \bar\cL}{k} W_0 + \frac{4\pi^2\bar\cL^2}{k^2}e^{-2\rho}W_{2}+\frac{4\pi \bar\cW}{k} e^{-\rho}\tm _{-1} \right) dx^{-} + 2\xi J d\phi - L_0 d\rho,  
\end{align}
where $L_0 = \tp_0 + \tp_-$ and $\sigma=(3/4)^2$. These differ from the connection of  \cite{Gutperle:2011kf} by a  gauge transformation 
$U=e^{\rho \tp_0}$ and also contains a gauge field in the $u(1)$. 

We shall consider the supersymmetry of the black hole with $\cW=0$ and   $\mu=0$ but
$\bar\cW\neq 0$ and  $\bar\mu \neq 0$.  Imposing this condition is analogous to imposing 
the extremality condition for the case of the BTZ black hole. 
The equation for the Killing spinor is  given by 
\begin{equation}\label{cov3}
\cD_\mu \lambda \equiv \pd_\mu  \lambda + [A_\mu , \lambda] =0, 
\end{equation}
and $\lambda$ is  expanded as 
\begin{equation}
\lambda \equiv \sum_{r=-1/2}^{1/2}  \e ^{r}G^+_{r} + \sum_{r=-1/2}^{1/2}  \be ^{r}G^-_{r} +\sum_{r=-3/2}^{3/2}  \l ^{r}U^+_{r} + \sum_{r=-3/2}^{3/2}  \tl ^{r}U^-_{r} . 
\end{equation}
Written   in matrix form  the equation given in  (\ref{cov3}) reads
\begin{align}
\pd _\rho   \begin{pmatrix}
\e^{-1/2} \\
\e^{1/2} \\
\l^{-3/2} \\
\l^{-1/2} \\
\l^{1/2} \\
\l^{3/2}
\end{pmatrix} +  \begin{pmatrix}
\tfrac{1}{2} &0 &0 &0 &0 &0 \\
0 &-\tfrac{1}{2}  &0 &0 &0 &0 \\
0 &0 &\tfrac{3}{2}  &0 &0 &0  \\
0 &0 &0  &\frac{1}{2} &0 &0 \\
0 &0 &0 &0  &-\frac{1}{2}  &0 \\
0 &0 &0 &0  &0  &-\tfrac{3}{2}
\end{pmatrix}  \begin{pmatrix}
\e^{-1/2} \\
\e^{1/2} \\
\l^{-3/2} \\
\l^{-1/2} \\
\l^{1/2} \\
\l^{3/2}
\end{pmatrix} =0, 
\end{align}

\begin{align}
\pd _+   \begin{pmatrix}
\e^{-1/2} \\
\e^{1/2} \\
\l^{-3/2} \\
\l^{-1/2} \\
\l^{1/2} \\
\l^{3/2}
\end{pmatrix} +  \left(
\begin{array}{cccccc}
 \xi  & \frac{8 \cL \pi }{3 k} & i & 0 & -\frac{2 i \cL \pi }{3 k} & 0 \\
 \frac{4}{3} & \xi  & 0 & \frac{i}{3} & 0 & -\frac{2 i \cL \pi }{k} \\
 -\frac{4 i \cL \pi }{3 k} & 0 & \xi  & \frac{4 \cL \pi }{3 k} & 0 & 0 \\
 0 & -\frac{4 i \cL \pi }{3 k} & 2 & \xi  & \frac{8 \cL \pi }{3 k} & 0 \\
 \frac{2 i}{3} & 0 & 0 & \frac{4}{3} & \xi  & \frac{4 \cL \pi }{k} \\
 0 & \frac{2 i}{3} & 0 & 0 & \frac{2}{3} & \xi 
\end{array}
\right)
 \begin{pmatrix}
\e^{-1/2} \\
\e^{1/2} \\
\l^{-3/2} \\
\l^{-1/2} \\
\l^{1/2} \\
\l^{3/2}
\end{pmatrix} =0. 
\end{align}
\begin{align}
\pd _-   \begin{pmatrix}
\e^{-1/2} \\
\e^{1/2} \\
\l^{-3/2} \\
\l^{-1/2} \\
\l^{1/2} \\
\l^{3/2} 
\end{pmatrix} + \left(
\begin{array}{cccccc}
 -\xi  & 0 & 0 & 0 & 0 & 0 \\
 0 & -\xi  & 0  & 0 & 0& 0 \\
 0 & 0 &-\xi  & 0 & 0& 0 \\
 0 & 0 & 0 & -\xi  & 0 &0 \\
 0 &0 & 0 & 0 &-\xi  & 0 \\
 0  & 0 & 0 &0  & 0 & -\xi 
\end{array}
\right)
  \begin{pmatrix}
\e^{-1/2} \\
\e^{1/2} \\
\l^{-3/2} \\
\l^{-1/2} \\
\l^{1/2} \\
\l^{3/2}
\end{pmatrix} =0. 
\end{align}
The solutions to these equations are given by
\begin{align}
\begin{pmatrix}
\e^{-1/2} \\
\e^{1/2} \\
\l^{-3/2} \\
\l^{-1/2} \\
\l^{1/2} \\
\l^{3/2}
\end{pmatrix} = \mathcal{R} (\rho) f_+(x_+) f_-(x_-), 
\end{align}
where, $\R (\rho)$ is defined in  (\ref{r-dep}) and 
\begin{align} \label{hdspin}
f_+ (x_+) &=  e^{-(-2\sqrt{\frac{2\pi \cL}{k}}+\xi)x_+} (c_1 \mathsf{y_1} +c_2 \mathsf{y_2})  +   e^{-(2\sqrt{\frac{2\pi \cL}{k}}+\xi)x_+} (c_3 \mathsf{y_3} +c_4 \mathsf{y_4})\nn \\ &\hspace{5cm} +  e^{-\xi x_+}  (c_4 \mathsf{y_4} +c_5 \mathsf{y_5}) . 
\end{align}
$\mathsf{y_i}$ are the eigenvectors of the matrix that appears in the + component of the Killing spinor equation.  As usual the $x_-$ dependence is  given by 
\begin{align}
f_- (x_-) = e^{\xi  x_-} . 
\end{align}
The value of the $u(1)$ field for which we get the proper periodicity of the spinor is
\begin{equation}\label{hdkill}
\xi = \pm \sqrt{\frac{2\pi \cL}{k}} ,   \qquad \text{or} \qquad \xi=i\frac{n}{2} .
\end{equation}
From degeneracy of the eigenvalues in (\ref{hdspin}) we see that in general we can have
two Killing spinors for a given $\xi$ satisfying any one of the conditions in (\ref{hdkill}).

\subsubsection*{A new higher spin black hole }

We shall now try to generalize the gauge connection (\ref{w3bh-c}) by including terms which involve the $sl(2)$ corresponding to the 
$\tm_\pm, \tm_0$ generators. This solution is same as the one 
given in  (\ref{w3bh-c}) 
but with the $sl(2)$ connections of BTZ the black hole added to it. It may thus admit a notion of the horizon.  The connection is given as follows and we have verified that it  obeys the flatness conditions. 
\begin{align}\label{w3bh-c2}
A=& \left( e^\rho  \tm _1 - \frac{2\pi}{k}\cL _1 e^{-\rho}\tm _{-1} + \frac{\pi}{2k\sigma} \cW e^{-2\rho}W_{-2} +  e^\rho  \tp _1 - \frac{2\pi}{k}\cL _2 e^{-\rho}\tp _{-1}    \right) dx^+ \nn \\
     & +\mu\left( e^{2\rho} W_{2} -\frac{4\pi \cL _1}{k} W_0 + \frac{4\pi^2\cL _1^2}{k^2}e^{-2\rho}W_{-2}+\frac{4\pi \cW}{k} e^{-\rho}\tm _{-1} \right) dx^{-} + 2\xi J d\phi,  \nn \\
    &    +(\tm _0 + \tp _0)d\rho \\
\bar A=& - \left( e^\rho  \tm_1 - \frac{2\pi}{k} \bar\cL _1 e^{-\rho}\tm_{-1} + \frac{\pi}{2k\sigma} \bar\cW e^{-2\rho}W_{-2} +  e^\rho  \tp _1 - \frac{2\pi}{k}\bar \cL _2 e^{-\rho}\tp _{-1}     \right) dx^+ \nn \\
     & -\bar\mu\left( e^{2\rho} W_{2} -\frac{4\pi \bar\cL _1}{k} W_0 + \frac{4\pi^2\bar\cL _1^2}{k^2}e^{-2\rho}W_{-2}+\frac{4\pi \bar\cW}{k} e^{-\rho}\tm_{-1} \right) dx^{-} + 2\xi Jd\phi,   \nn \\
    &    -(\tm _0 + \tp _0)d\rho.  
\end{align}
with $\sigma=(3/4)^2$. The metric due to the above gauge connections is 
\begin{align}
ds^2 &= d\rho^2 -3 \left( \mu e^{2\rho}dx^- + \frac{16\pi}{18k}{\bar \cW} + \frac{4\pi^2}{k^2}{\bar \mu} {\bar \cL _1}^2 e^{-2\rho} dx^+ \right)\nn \\
& \hspace{1.45cm} \times \left({\bar \mu} e^{2\rho}dx^- + \frac{16\pi}{18k}{ \cW} + \frac{4\pi^2}{k^2}{\mu} {\cL _1}^2 e^{-2\rho} dx^+ \right)\nn \\
& \quad-\frac{4}{3} \left( e^\rho dx^+ - \frac{2\pi}{k} {\bar \cL}_1 e^\rho dx^- +\frac{4\pi}{k}{\bar \mu}{\bar \cW}e^{-\rho}dx^+   \right)  \left( e^\rho dx^+ - \frac{2\pi}{k} {\cL}_1 e^\rho dx^- +\frac{4\pi}{k}{\mu}{\cW}e^{-\rho}dx^+   \right) \nn \\
& \quad -\frac{1}{4} \left( \frac{4\pi}{k}\right)^2 (\mu\cL _1 dx^- + {\bar\mu\cL }_1 dx^+)^2
 - \frac{2\pi}{3k}\left( \cL _2 (dx^+)^2 +{ \bar \cL}_2 (dx^-)^2   \right) \nn \\
&\quad + \frac{1}{3} \left( e^{2\rho} + \left(\frac{2\pi}{k}\right)^2 \cL _2{\bar \cL}_2 e^{-2\rho} \right)dx^+dx^-
\end{align}
We shall again consider the supersymmetry of the black hole with $\cW=0$ and $\mu=0$. The equation for the Killing spinor is given by 
\begin{equation}\label{cov23}
\cD_\mu \lambda \equiv \pd_\mu  \lambda + [A_\mu , \lambda] =0, 
\end{equation}
where $\lambda$ is given by
\begin{equation}
\lambda \equiv \sum_{r=-1/2}^{1/2}  \e ^{r}G^+_{r} + \sum_{r=-1/2}^{1/2}  \be ^{r}G^-_{r} +\sum_{r=-3/2}^{3/2}  \l ^{r}U^+_{r} + \sum_{r=-3/2}^{3/2}  \tl ^{r}U^-_{r} . 
\end{equation}
Written  in matrix  form  the equation in  (\ref{cov23}) reads
\begin{align}
\pd _\rho   \begin{pmatrix}
\e^{-1/2} \\
\e^{1/2} \\
\l^{-3/2} \\
\l^{-1/2} \\
\l^{1/2} \\
\l^{3/2}
\end{pmatrix} + \begin{pmatrix}
\tfrac{1}{2} &0 &0 &0 &0 &0 \\
0 &-\tfrac{1}{2}  &0 &0 &0 &0 \\
0 &0 &\frac{3}{2}  &0 &0 &0  \\
0 &0 &0  &\tfrac{1}{2} &0 &0 \\
0 &0 &0 &0  &-\tfrac{1}{2}  &0 \\
0 &0 &0 &0  &0  &-\tfrac{3}{2}
\end{pmatrix}  \begin{pmatrix}
\e^{-1/2} \\
\e^{1/2} \\
\l^{-3/2} \\
\l^{-1/2} \\
\l^{1/2} \\
\l^{3/2}
\end{pmatrix} =0, 
\end{align}
\begin{align}
&\pd _+   \begin{pmatrix}
\e^{-1/2} \\
\e^{1/2} \\
\l^{-3/2} \\
\l^{-1/2} \\
\l^{1/2} \\
\l^{3/2}
\end{pmatrix} +  \\
&\qquad \left(
\begin{array}{cccccc}
 \xi  & \frac{2\pi(4 \cL_1 -\cL_2) }{3 c}  & 0 & 0 & -\frac{2 i \pi  \left(\cL_1-\cL_2\right)}{3 c} & 0 \\
 1 & \xi  & 0 & 0 & 0 & -\frac{2 i \pi  \left(\cL_1-\cL_2\right)}{c} \\
 -\frac{4 i \pi  \left(\cL_1-\cL_2\right)}{3 c} & 0 & \xi  & \frac{2 \pi  \left(2 \cL_1+\cL_2\right)}{3 c} & 0 & 0 \\
 0 & -\frac{4 i \pi  \left(\cL_1-\cL_2\right)}{3 c} & 3 & \xi  & \frac{4 \pi  \left(2 \cL_1+\cL_2\right)}{3 c} & 0 \\
 0 & 0 & 0 & 2 & \xi  & \frac{2 \pi  \left(2 \cL_1+\cL_2\right)}{c} \\
 0 & 0 & 0 & 0 & 1 & \xi 
\end{array}
\right)
\begin{pmatrix}
\e^{-1/2} \\
\e^{1/2} \\
\l^{-3/2} \\
\l^{-1/2} \\
\l^{1/2} \\
\l^{3/2}
\end{pmatrix} =0, \nn
\end{align}
\begin{align}
\pd _-   \begin{pmatrix}
\e^{-1/2} \\
\e^{1/2} \\
\l^{-3/2} \\
\l^{-1/2} \\
\l^{1/2} \\
\l^{3/2} 
\end{pmatrix} + \left(
\begin{array}{cccccc}
 -\xi  & 0 & 0 & 0 & 0 & 0 \\
 0 & -\xi  & 0  & 0 & 0& 0 \\
 0 & 0 &-\xi  & 0 & 0& 0 \\
 0 & 0 & 0 & -\xi  & 0 &0 \\
 0 &0 & 0 & 0 &-\xi  & 0 \\
 0  & 0 & 0 &0  & 0 & -\xi 
\end{array}\right)
  \begin{pmatrix}
\e^{-1/2} \\
\e^{1/2} \\
\l^{-3/2} \\
\l^{-1/2} \\
\l^{1/2} \\
\l^{3/2}
\end{pmatrix} =0. 
\end{align}
The solutions to these equations are given by 
\begin{align}
\begin{pmatrix}
\e^{-1/2} \\
\e^{1/2} \\
\l^{-3/2} \\
\l^{-1/2} \\
\l^{1/2} \\
\l^{3/2}
\end{pmatrix} = \mathcal{R} (\rho) f_+(x_+) f_-(x_-), 
\end{align}
where, $\cal{R}(\rho)$ is the square matrix in (\ref{r-dep}) which contains the $\rho$-dependence. The $x_+$ and $x_-$ dependent pieces are as follows
\begin{align}
f_+ (x_+) =& c_1 e^{-(-\sqrt{\frac{2\pi \cL _2}{k}}+\xi)x_+}  \mathsf{y_1}  +   c_2 e^{-(\sqrt{\frac{2\pi \cL _2}{k}}+\xi)x_+}  \mathsf{y_2} + c_3 e^{-(-\sqrt{\frac{2\pi \cL _2}{k}}     +  2\sqrt{\frac{2\pi \cL _1}{k}}  +\xi)x_+} \mathsf{y_3}\nn \\
& \quad + c_4 e^{-(\sqrt{\frac{2\pi \cL _2}{k}}  - 2\sqrt{\frac{2\pi \cL _1}{k}}  +\xi)x_+}\mathsf{y_4} + c_5 e^{-(-\sqrt{\frac{2\pi \cL _2}{k}}     -  2\sqrt{\frac{2\pi \cL _1}{k}}  +\xi)x_+} \mathsf{y_5}\nn \\
&\quad +c_6 e^{-(\sqrt{\frac{2\pi \cL _2}{k}}   +  2\sqrt{\frac{2\pi \cL _1}{k}}  +\xi)x_+} \mathsf{y_6}, 
\end{align}
$\mathsf{y_i}$ are the eigenvectors of the matrix that appears in the + component of the Killing spinor equation. 
\begin{align}
f_- (x_-) = e^{\xi  x_-}.   
\end{align}
The value of the $u(1)$ field for which we obtain  periodic Killing spinors  is given by 
\begin{equation}
\xi = \pm \left( \sqrt{\frac{2\pi \cL_1}{k}} \pm \frac{1}{2} \sqrt{\frac{2\pi \cL_2}{k}}  \right) \qquad \text{or,} \qquad \xi=  \pm \frac{1}{2} \sqrt{\frac{2\pi \cL_2}{k}} \  .
\end{equation}
\\
Thus generically the solution admits a single Killing spinor. 

For the case of the black holes in this paper we have explicitly solved the Killing spinor components of $G^+_r$ and $U^+_r$. The same method can be employed to solve for the components of the $G^-_r$ and $U^-_r$ generators as well.

\subsection{Summary of the solutions and their supersymmetry } \label{table}

We now summarize the results for the Killing spinors for the various classical 
solutions of the $sl(3|2)$ Chern-Simons theory considered in this paper. 
The generic 
number of Killing spinors and their periodicity condition
listed in this table correspond to the $G^+$ and $U^-$ generators
of the theory.

\newpage
\begin{center}
{\bf Table 1}
\end{center}

\begin{footnotesize}
 \begin{tabular}{|c |c|c|} 
 \hline 
\textbf{Background }&\textbf{Killing spinor condition }&\textbf{Number of Killing spinors }\\
\hline \hline
\shortstack{\text{BTZ black hole }\\ {in} $sl(2)$ with $M=J$}  & \shortstack{ Periodic Killing spinors } &3 \\  \hline 
\shortstack{\text{BTZ black hole}\\ with $M=J$}  & \shortstack{ Periodic Killing spinors } &2 \\ \hline    
\shortstack{\\ \text{Higher spin} \text{black hole}\\  of Gutperle et al. \\  with $\mathcal{W}=\mu=0$} &\shortstack{ $\xi = \pm \sqrt{\frac{2\pi \cL}{k}}$ {or} \\ $ \xi=i\frac{n}{2} $ } &2  \\ \hline  
\shortstack{\text{New R-charged}\\ \text{higher spin black hole}  \\ with $ \mathcal{W}=\mu=0$} &\shortstack{
$ \xi = \pm \left( \sqrt{\frac{2\pi \cL_1}{k}} \pm \frac{1}{2} \sqrt{\frac{2\pi \cL_2}{k}}  \right) $ or \\ $ \xi=  \pm \frac{1}{2} \sqrt{\frac{2\pi \cL_2}{k}}   $ 
}   &1    \\ \hline   

\hline
\shortstack{\text{$AdS_3$} {in} $sl(2)$}  & \shortstack{ Anti-periodic Killing spinors } &6  \\  \hline
{\text{$AdS_3$}}  & \shortstack{ Anti-periodic Killing spinors } &6 \\ \hline 
\shortstack{Higher spin \\ conical defect} &\shortstack{$ 2\xi = \pm i\sqrt{\alpha} \delta +in $ \\ $2\xi = \pm i\sqrt{\alpha} \left(   \delta \pm 2 (\beta ^2 - (\tfrac{3}{4}\eta)^2 )^{1/2}  \right) +in  $}   &1  \\ \hline
\shortstack{\text{Conical defects}\\ \text{in $sl(2)$}} & $\xi=\pm i \frac{\sqrt{\gamma}}{4} + in$  &3 \\ \hline  
\shortstack{\text{Conical defects}\\ \text{in gravitational $sl(2)$}} & $\xi= \pm i \frac{\sqrt{\gamma}}{4}+in$  \text{or} $ \xi= \pm 3i \frac{\sqrt{\gamma}}{4}+in.$ &1  \\ \hline   
\end{tabular}

\end{footnotesize}

\section{Supersymmetry and holonomy}

\def\bi{\bar{\imath}}
\def\bj{\bar{\jmath}}

In the previous section we have solved for the conditions under which 
the background solution admits periodic Killing spinors. 
This was a tedious but straight forward exercise. 
Since the  Chern-Simons action is independent of the metric on the manifold it 
must be possible to state these conditions in terms of gauge independent 
and well defined physical observables. 
In this section we show that the periodicity conditions for the Killing spinor 
  can be  written in terms of a condition on the
  eigenvalues of  holonomy of the  background gauge connection around the 
angular $\phi$ direction.  This invariant characterization of supersymmetry 
in higher spin theories in 3 dimension is the central result of this work. 
We state the condition for a general gauge connection belonging to the 
$sl(N|N-1)$ superalgebra. 
We show that whenever  the holonomy of the $u(1)$ part of the connection along with eigenvalues of the rest of the  background holonomy  weighted with
the odd roots of the superalgebra becomes trivial then the Killing spinor is periodic. 
This condition is given in equation (\ref{periodf}). 
We then explicitly verify that this  condition reproduces the equations
(\ref{spinor-cond-1})  and (\ref{spinor-cond-2}) we find for the  higher 
spin conical defects in the $sl(3|2)$ algebra. We have also verified that the
holonomy condition reproduces the periodicity of Killing spinor for the black holes
considered in the $sl(3|2)$ theory. 
We then proceed to combine the supersymmetry requirement along with the 
requirement that the holonomy is smooth to show that for $N\geq 4$ the 
$sl(N|N-1)$ theory admits smooth and supersymmetric conical 
defects.

\subsection{Killing spinor periodicity as a holonomy}\label{spinper}

The equation for the covariantly constant  Killing spinor satifies the equation given by 
\begin{equation}\label{kill-1}
\cD_\mu \epsilon \equiv \pd _\mu \epsilon + [A_\mu , \epsilon ] = 0. 
\end{equation}
Here  $\epsilon = \epsilon^i G_i$ is a linear combination of the fermionic generators.
$A_\mu = A_\mu^a T_a$ are the connection one forms valued in the 
bosonic part of the algebra. 
It is convenient to choose  the fermionic generators in the Cartan-Weyl basis
of the super algebra. For definiteness we can work with the super group $sl(N|N-1)$
but the discussion can be easily generalized to any super algebra. 
In the Cartan-Weyl basis, the generators  satisfy the following conditions: let $H_r$ be the 
Cartan's of the superalgebra  and $J$ be the $U(1)$. Then we have the commutation relations
\begin{equation} \label{cwc}
[H_r, G_i] = \alpha^{r}_i G_i, \qquad [J, G_i] = \pm G_i, 
\end{equation}
were $\alpha^r_i$ is the $r$th component of the odd root $\alpha_i$. 
As mentioned in  section 2 we see that the integrability condition for the 
Killing spinor equation is satisfied since the background gauge field 
satisfies the equation of motion. We can therefore solve the equation in 
(\ref{kill-1}) formally by writing the solution as 
\begin{equation}\label{kilsol}
 \epsilon (x)  ={\cal P}(  e^{ \int_{x_0}^x A_\mu dx^\mu} )  \hat \epsilon (x_0)
{\cal P}(  e^{ -\int_{x_0}^x A_\mu dx^\mu} ) , 
\end{equation}
where $x_0$ is a base point and $\hat \epsilon(x_0)$ is a constant spinor and ${\cal P}$ 
refers to the path ordered exponential. 
To determine whether the spinor is periodic we can consider $x= (\rho, t, 2\pi)$ and 
$x_0  = (\rho, t , 0)$ and the integral is along the  constant time circle in the angular 
direction. For all the solutions considered in this paper, the holonomy along this circle reduces
to the form
\begin{equation}
{\rm Hol}_\phi (A) = {\cal P} \exp( \oint A_\mu dx^\mu) = 
 b^{-1}(\rho)  \exp \left( \oint a_\phi d\phi \right)  b(\rho) , 
\end{equation}
where $b(\rho)$ is the matrix which contains the $\rho$ dependence. 
The connection $a_\phi$ is constant and can be easily integrated. 
Since it is a sum of  the bosonic
generators we can write it as 
\begin{equation}\label{conhol}
\exp \left( \oint a_\phi d\phi \right)  = S^{-1} \exp ( 2\pi  ( \lambda ^r H_r  +  2 \xi J))  S, 
\end{equation}
where $S$ is the similarity transformation which brings the constant holonomy in the diagonal form. 
Now substituting the equation (\ref{conhol}) in the solution of the Killing spinor 
given in (\ref{kilsol}) we find the periodicity of the spinor is determined by 
\begin{equation} \label{pkill}
\epsilon ( \rho, t, 2\pi) = b^{-1} S^{-1} e^{ 2\pi  (  \lambda ^r H_r  +   \xi J) }S
 b (\rho) \hat \epsilon(\rho, t , 0) b^{-1} (\rho)  S^{-1} e^{ -2\pi ( \lambda ^r H_r  + 2 \xi J)} S b . 
\end{equation}
Since the Cartan-Weyl basis for fermionic generators $G_i$  is complete we
have the relation
\begin{equation} \label{simt}
S b(\rho) \hat\epsilon(\rho, t,  0) b^{-1} (\rho)  S^{-1}= \epsilon (\rho, t, 0 ) = \tilde \epsilon^i(\rho, t, 0 ) G_i. 
\end{equation}
From the commutation relations given in (\ref{cwc}) we find
\begin{equation} \label{phase}
e^{2\pi ( \lambda ^r H_r  +   \xi J) }G^i e^{ -  2\pi ( \lambda ^r H_r  + 2 \xi J)  }
= e^{ 2\pi (   \lambda^r \alpha_i^r \pm2  \xi)  } G_i.
\end{equation}
Now substitute equations (\ref{simt}) and (\ref{phase}) into the periodicity
constraint  for the Killing spinor given in (\ref{pkill}). Let us 
 consider the case in which say any one of the $\tilde \epsilon^i$ is turned on and the rest 
set to zero.  
Then  we see that the spinor
with $\tilde \epsilon^i$ along the generator $G^i$ is periodic if
the following condition is true
\begin{equation} \label{periodf}
 \lambda^r \alpha_i^r \pm  2\xi   = i n. 
 \end{equation}
 where $n$ is any integer and $r$ is summed over the Cartan directions other than the 
 $U(1)$.  Recall that $\lambda^r$ are the eigenvalues of the holonomy of the 
background connection,  $\alpha_i^r$ are the odd roots of the Cartan generator
$H_r$ and $\xi$ is the value of the $U(1)$ field. 
 Note that the sign $\pm$ depends on the sign of the commutation relation $[J, G^i] = \pm G^i$. 
 Thus we find that the periodicity property of the Killing spinor along the $\phi$ direction 
 can be generally stated in terms of product of the eigenvalues of the 
holonomies of the background connection 
 with the  odd roots of the super algebra.  The number of supersymmetries preserved 
can also be found easily by checking how many among all the fermionic directions
labelled by $i$ satisfy the condition (\ref{periodf}). 
We have thus shown that  the supersymmetry condition on any background can be 
written in terms of gauge invariant and physically independent observables.

\subsubsection*{A test of  the supersymmetry condition}

We will now verify  the general equation for the 
periodicity of the Killing spinor  derived  in (\ref{periodf}) for the 
specific situation of higher spin conical defects in the $sl(3|2)$ theory. 
From the  gauge connection in (\ref{gauge-cone}) we obtain 
\begin{equation}
a_{\phi} = \delta_{-1}T^+_{-1} + \delta_{1}T^+_{1}  +  \beta_{-1}T^-_{-1} + \beta_{1}T^-_{1}  +  \eta_{-1}W_{-1} + \eta_{1}W_{1}  + 2 \xi J. 
\end{equation}
We now use  representation of the matrices for $sl(3)$  given in \cite{Gutperle:2011kf}
 with $\sigma=(\tfrac{3}{4})^2$ and the following 
representation for $sl(2)$  in terms of the Pauli matrices  
\begin{equation}
T^+_1=\tfrac{1}{2}(\sigma_1-i \sigma_2), \qquad  T^+_{-1}=\tfrac{1}{2}(\sigma_1+i \sigma_2),
\qquad
 T^+_0=\tfrac{1}{2}\sigma_3. 
\end{equation}
Then the eigenvalues of the  $sl(3)\oplus sl(2)$ part of the  matrix $a_\phi$  along with the $u(1)$ part 
is given by 
\begin{equation}\label{diagg}
Sa_\phi  S^{-1} = \text{Diag} \left[ 2i \sqrt{\alpha\left( \beta^2 - (\tfrac{3}{4} \eta)^2 \right) }, 0 ,
-2i \sqrt{\alpha\left( \beta^2 - (\tfrac{3}{4} \eta)^2 \right) }, 
i \sqrt{\alpha }\delta , - i \sqrt{\alpha} \delta \right]
+ 2 \xi J. 
\end{equation}
We will now write this as a linear combination of the Cartan generators of $sl(3|2)$. From the 
appendix which lists the generators of $sl(N|N-1)$,  we find that (\ref{diagg}) can be written as 
\begin{equation}
 Sa_\phi  S^{-1} =  2i \sqrt{\alpha( \beta^2 -(\tfrac{3}{4} \eta)^2 ) } ( H_1 + H_2)  + 
i \sqrt{\alpha } \delta  H_{\bar 4} + 2\xi J, 
\end{equation}
 where the Cartan matrices are given by 
\begin{align}
H_1 = \begin{pmatrix}
1 & & & & \\
 &-1 & & & \\
 &  &0 & & \\
 &  &  &0 & \\
 & & & &0 
\end{pmatrix} \qquad &H_2 = \begin{pmatrix}
0 & & & & \\
 &1 & & & \\
 &  &-1 & & \\
 &  &  &0 & \\
 & & & &0 
\end{pmatrix}  \qquad
 H_{\bar{4}} = \begin{pmatrix}
0 & & & & \\
 &0 & & & \\
 &  &0 & & \\
 &  &  &1 & \\
 & & & &-1 
\end{pmatrix}. 
\end{align}
In this representation  the $U(1)$ generator $J$ is given by 
\begin{equation} \label{defj}
 J = \begin{pmatrix}
-2 & & & & \\
 &-2 & & & \\
 &  &-2 & & \\
 &  &  &-3 & \\
 & & & &-3 
\end{pmatrix}. 
\end{equation}
We  now need the odd roots of the supercharges with $J$ charge 1. 
In the Cartan-Weyl basis these are given by 6 matrices $E_{\bi , k}$ with 
$\bi={\bar 4,\bar  5}$ and $k =1, 2, 3$.  They correspond to the $6$  generators 
$G^+_{\pm1/2}, U^+_{\pm 1/2}, U^+_{\pm 3/2}  $. 
Evaluating the commutation relations explicitly using the matrix representation 
given in the appendix we find the following roots
\begin{align}\label{eroot}
[H_1 + H_2 ,E_{\bi 1}]=& -E_{\bi 1}, \qquad
[H_1 + H_2 ,E_{\bi 2}] = 0, \qquad  
[H_1 + H_2 ,E_{\bi 3}] = E_{\bi 3}, \nonumber  \\ 
&[H_{\bar 4} ,E_{\bar 4 k }]= E_{\bar 4 i } , \qquad
[H_{\bar 4} ,E_{\bar 5 k }] = -E_{\bar{5} k} . 
\end{align}
We now have all the information to derive the supersymmetric conditions given in 
(\ref{periodf}). Consider the supercharge  $E_{\bar 4 1}$, using the holonomy of the 
background given in (\ref{diagg}) and the roots from (\ref{eroot}) we find the following 
condition
\begin{equation}
- 2i \sqrt{\alpha( \beta^2 - (\tfrac{3}{4} \eta)^2) } + i \sqrt{\alpha } \delta   + 2\xi = in.
\end{equation}
We see that this matches with one of equations in (\ref{spinor-cond-2}). Now consider the 
supercharge $E_{\bar 4 2}$, again using (\ref{diagg}) and (\ref{eroot})  we obtain
\begin{equation}
 i \sqrt{\alpha } \delta   + 2\xi = in. 
 \end{equation}
 This coincides with one of the equations in (\ref{spinor-cond-1}). Repeating this 
 explicitly for all the remaining supercharges we obtain the $6$ conditions 
 in (\ref{spinor-cond-1}) and (\ref{spinor-cond-2}). 
 We have also verified that the supersymmetry condition (\ref{periodf}) reproduces 
 the conditions for the periodicity of the Killing spinors for the case of black holes 
 in the $sl(3|2)$ theory studied in this paper.

\subsection{Smooth holonomy and supersymmetry}

Smooth conical defects have played a central  role in tests of 
the minimal model/higher spin duality.  They are dual to the primaries 
of the ${\cal W}_N$ minimal model after a suitable analytical continuation 
\cite{Castro:2011iw}. 
A Kazama-Suzuki  supersymmetric minimal model  was proposed to be 
dual to  the large $N$ limit of the $sl(N|N-1)$ higher 
spin theories studied in this paper \cite{Creutzig:2011fe}. 
Thus we expect smooth conical defects to be dual to 
primaries of the supersymmetric minimal model. 
However in a supersymmetric theory there are special primaries 
called chiral primaries which preserve supersymmetry. They are protected 
against quantum 
corrections and can be used as probes to test the minimal model/higher spin 
duality.  Thus smooth supersymmetric conical defects of the $sl(N|N-1)$ theory
 are expected to  be dual to 
the chiral primaries of the Kazama-Suzuki minimal model after an analytic
continuation to infinite $N$ or finite central charge \footnote{The gravitational 
higher spin theory we are studying is classical and therefore has large central charge}. 
Note that all the conical defects 
considered in this paper are metrically singular as seen from the 
metric written in equation (\ref{conical-metric2}). However since the circle  around the angular direction
$\phi$, is contractable a gauge invariant method to decide when the solution 
is smooth is to consider the 
holonomy of  the Chern-Simons connection around this circle \cite{Castro:2011iw}. 
A solution is smooth if this holonomy is trivial.
With this motivation  we study  the conditions under which 
a conical defect is both smooth as well as supersymmetric.
We first begin with the $sl(3|2)$ theory and show that 
it does not admit smooth superymmetric conical defects. 
The supersymmetric defects are singular in this case,  that is they 
do not admit a smooth holonomy. We then study supersymmetry and smoothness for conical 
defects in $sl(N|N-1)$ theories for $N\geq 4$. 
For these theories it is shown that there are smooth and supersymmetric
conical defects.

\subsubsection*{ Smoothness and supersymmetry: $sl(3|2)$}

Let us now focus on the  the $sl(3|2)$ theory and investigate if the theory 
admits smooth supersymmetric conical defects. 
As  discussed above we first 
 demand  that the   holonomy  around the 
 angular direction is trivial. This  leads to the 
 following conditions on $\delta, \ \beta$ and $\eta$. 
\begin{eqnarray}\label{holc}
&& \sqrt{\alpha}\delta = \pm  \frac{m}{2} , \nn \\
&&2\sqrt{\alpha} (\beta ^2 - (\tfrac{3}{4}\eta)^2 )^{1/2}=  \pm p \label{int-con-1},  \\
&&-2i\xi =  \pm q . \nn 
\end{eqnarray}
where $m, \ p, \ q \in \mathbb{Z} $. Note that the values of $\sqrt{\alpha} \delta$ are quantized 
in half integers because the center of $SL(2)$ is $\mathbb{Z}_2$ valued. 
Substituting  the periodicity conditions of the Killing spinors 
given in (\ref{spinor-cond-1}) and (\ref{spinor-cond-2}) we obtain the following conditions
\begin{align}\label{inte-cond}
q \mp \frac{m}{2} = n, \qquad q \mp ( \frac{m}{2} \pm  p ) = n . 
\end{align}

Let us now examine if any of the conical defects in the $sl( 3|2)$ theory satisfy the 
requirement that they lie in the domain  pointed out in \cite{Castro:2011iw}
\begin{equation}\label{bound}
-\frac{c}{24} < L_0 < 0, 
\end{equation}
where $c$ is the central charge of the theory which can be written in terms of the cosmological 
constant.  This restriction comes from the fact that we need these solutions 
to have mass above the $AdS_3$ vaccumm and below the zero mass BTZ. 
For the $sl(3|2)$ theory the value $L_0$ in terms of the holonomy is given by 
\begin{equation} \label{vall}
L_0 = \frac{c}{24 \epsilon_{(3|2)}} \left( {\rm str} ( a_{\phi}^2)  \right) 
\end{equation}
Note that by defining $L_0$ as given in (\ref{vall}) the shift in energy  due to the presence of 
 the presence of the $U(1)$ field
\cite{Henneaux:1999ib, Maldacena:2000dr} is accounted for. 
Substituting the values of the holonomy from (\ref{holc}) in (\ref{vall}) we obtain the 
following bound that the integers $p, q, m $ must satisfy
\begin{equation}
0< p^2 - \left( \frac{m}{2}\right)^2 - 6  q^2 <\frac{3}{4}. 
\end{equation}
The factor of $6$ occurs on taking the super trace of $J^2$. Thsi can be 
seen by using  the definition of $J$ given in (\ref{defj}). 
On substituting for $m/2$ from the  supersymmetric holonomy conditions given in 
(\ref{inte-cond}) we obtain the following bounds
\begin{equation}
0< p^2 - ( q -n)^2 - 6 q^2 < 3/4, \qquad
0< p^2 - ( q -n \mp p)^2  -6 q^2 <3/4. 
\end{equation}
It is clear that any of the above bounds are satisfied since there is no integer between
$0$ and $3/4$. Thus there are no supersymmetric smooth conical defects  in the $sl(3|2)$.

\subsubsection*{Smoothness and supersymmetry: $sl(N|N-1), N\geq 5$}

We will now look at the $sl(N|N-1)$ Chern-Simons theory for $N\geq 5$ and show
that the theory admits smooth conical defects. 
The case of $N=4$ will be treated later, the reason is that for $N\geq 5$, the Cartan
generators of $sl(N|N-1)$  can be stated in a simple form. 
We shall be using the algebra and the matrix representation of the generators given in Section 61 of 
\cite{Frappat:1996pb}. We have reviewed this  representation  in Appendix A. 
Following \cite{Castro:2011iw} we can write the gauge connections  for the  conical defects in 
the 
$sl(N|N-1)\oplus sl(N|N-1)$ theory as
\begin{equation}
A=b^{-1} a b + b^{-1}d b \ , \qquad \bar{A}=b^{-1} \bar{a} b + b^{-1}d b , 
\end{equation}
where $b=\text{exp}(\rho \hat L_0)$. 
$\hat L_0$ is the $sl(2)$ generator which is principally embedded in  $sl(N|N-1)$. 
Explicitly it is given by  the 
 diagonal $(2N-1) \times (2N-1)$ matrix whose diagonal  elements are given by
\begin{eqnarray} \label{hatl}
 2\hat L_0 =  {\rm Diag} (  N-1, N-3, \cdots -(N-3) , -(N-1), \\ \nonumber 
 N-2, N-4, \cdots - ( N-4), - ( N-2) ) . 
\end{eqnarray}
The connections $a$ and $\bar a$ are given by 
\begin{align} \label{sln-gc}
a &=\left( \sum_{k=1}^{N-1} B_k (a_k, \alpha a_k) +  \sum_{\bar{k}=\overline{N+1}}^{\overline{2N-2}} B_{\bar{k}} (a_{\bar{k}}, \alpha a_{\bar{k}}) \right) dx^+ + 2 \xi J,  \nn \\
\bar{a} & =-\left( \sum_{k=1}^{N-1} B_k (\gamma a_k, \tfrac{\gamma}{\alpha} a_k) +  \sum_{\bar{k}=\overline{N+1}}^{\overline{2N-2}} B_{\bar{k}}  (\gamma a_{\bar{k}}, \tfrac{\gamma}{\alpha} a_{\bar{k}}) \right) dx^- + 2 \xi J, 
\end{align}
where  `$B$-matrices' are defined as 
\begin{equation}
[B_K (x,y)]_{IJ} = x \delta _{I,K} \delta_{J,K+1} - y\delta_{I,K+1}\delta_{J,K}. 
\end{equation}
$I$, $J$ and $ K$ values run from $1,2, \dots , N, \overline{N+1},\overline{N+2},\dots,\overline{2N-1}$. 
The connection given in (\ref{sln-gc}) contains the conical defects found by 
\cite{Castro:2011iw}  in both the  algebras $sl(N)$ and $sl(N-1)$ together with the $u(1)$ field
turned on. 
We now have to diagonalize the connection $a_\phi$ to find the eigenvalues of the 
holonomy matrix. 
When the connection given in (\ref{sln-gc})  is brought to the diagonal form, 
 it can can written
as a linear combination of the Cartan generators $H_J$ of $sl(N|N-1)$ given in the
Appendix. For $N\geq 5$ we obtain  
\begin{equation} \label{lincomb}
Sa_\phi S^{-1}  =\begin{cases}
   i\underset{j \, \text{odd}}{ \sum_{j=1}^{N-1}} a_j H_j + i\underset{j \, \text{odd}} {\sum_{\bj =\overline{ N+1}} ^{\overline{2N-3}}}\, a_{\bj} H_{\bj} + 2\xi J , & \text{for $N$ even}.\\
     i\underset{j \, \text{odd}}{ \sum_{j=1}^{N-2} } a_j H_j + i\underset{j \, \text{even}}{ \sum_{\bj =\overline{ N+1}}^{\overline{2N-2}}}\, a_{\bj} H_{\bj} + 2 \xi J, & \text{for $N$ odd}.
  \end{cases}
\end{equation}
On imposing trivial holonomies for smoothness of the conical defect solutions, we have
\begin{align}\label{holo-int}
\text{For even $N$ } \ : \ & a_i=\frac{m_i}{2}, \ a_{\bi}= p_{\bi} \ , \ -2i\xi =q,  \nn \\
\text{For odd $N$   } \ : \ & a_i={r_i}\ , \ a_{\bi}=\frac{ s_{\bi} }{2} \ , \ -2i\xi =t.  
\end{align}
Here $p_{\bi},q,r_i,t \in \mathbb{Z} $ and $m_i$ and $s_{\bar i}$ take values
in the set of either even or odd integers for all $i$ and $\bar i$ respectively. 
The reason that for even $N$, $a_i$ takes values in the set of half integers is because
the group $SL(N)$ has a $\mathbb{Z}_2$ valued center. Similarly for odd
$N$ the group $SL(N-1)$ has a $\mathbb{Z}_2$ valued center which makes
$a_{\bi}$ takes values in the set of half integers. 

\def\bl{{\bar{l}}}

The next step is to 
 find out the roots of the fermionic generators, $E_{i\bj}$ with the linear combination of the Cartan matrices given in (\ref{lincomb}).  
A generic linear combination can be written as $\sum_k a_k H_k +\sum_\bl a_\bl H_\bl $. The commutator of this with a fermionic generator is 
\begin{align}\label{root1}
\left[i\sum_k a_k H_k +i\sum_\bl a_\bl H_\bl \; , E_{i\bj} \right] =i [(a_i-a_{i-1})-(a_{\bj}-a_{\overline{\jmath-1}})] E_{i\bj}. 
\end{align}
Here $a_0= a_{\bar 0} =0$ and these fermionic generators have $u(1)$ charge $+1$. 
Using these roots we can write out the periodicity condition for the Killing spinors
given in (\ref{periodf}).  We see that the conditions split to 
 four cases each for even and odd $N$. 
For even $N$
\begin{align}
\text{odd $i$ and odd $\bj$}&: \ i(a_i - a_{\bj}) + 2\xi = in _{i\bj}, \nn  \\
\text{odd $i$ and even $\bj$}& : \ i(a_i + a_{\overline{\jmath - 1}})+ 2\xi = in _{i\bj},  \nn \\
\text{even $i$ and odd $\bj$}&: \ -i(a_{i-1} + a_{\bj}) + 2\xi = in _{i\bj},  \nn \\
\text{even $i$ and even $\bj$}&: \ -i(a_{i-1} - a_{\overline{\jmath - 1}})+ 2\xi = in _{i\bj}. 
\end{align}
Whereas for odd $N$ we have
\begin{align}
\text{odd $i$ and odd $\bj$}&: \  i(a_i + a_{\overline{\jmath - 1}})+ 2\xi = in _{i\bj},  \nn \\
\text{odd $i$ and even $\bj$}& : \  i(a_i - a_{\bj})+ 2\xi = in _{i\bj}, \nn \\
\text{even $i$ and odd $\bj$}&: \ -i(a_{i-1} - a_{\overline{\jmath - 1}}) + 2\xi = in _{i\bj}, \nn \\
\text{even $i$ and even $\bj$}&: \  -i(a_{i-1} + a_{\bj})+ 2\xi = in _{i\bj}. 
\end{align}
where \( n_{i\bj}\in\mathbb{Z} \). 
Thus there are $N(N-1)$ such conditions which is equal to the 
number of fermionic generators with positive $u(1)$ charge. Substituting 
the quantization conditions of  (\ref{holo-int}) in the above equations  we obtain
the following constraints from  the periodicity of the Killing spinors. 
\begin{align}\label{sln-period}
\text{even $N$ : } \begin{cases}
\text{odd $i$ and odd $\bj$}&: \ \frac{m_i}{2}-p_{\bj} +q =n_{i\bj},  \\
\text{odd $i$ and even $\bj$}& : \ \frac{m_i}{2}-p_{\overline{\jmath - 1}} +q =n_{i\bj},  \\
\text{even $i$ and odd $\bj$}&: \ -\frac{m_{i-1}}{2}-p_{\bj} +q =n_{i\bj},  \\
\text{even $i$ and even $\bj$}&: \ -\frac{m_{i-1}}{2}+p_{{\overline{\jmath - 1}}} +q =n_{i\bj}.  
\end{cases} \\
\text{odd $N$ : } \begin{cases}
\text{odd $i$ and odd $\bj$}&: \    r_i - \frac{  s_{\overline{\jmath - 1}}}{2} +t =n_{i\bj},  \\
\text{odd $i$ and even $\bj$}& : \  r_i - \frac{  s_{\bj} }{2}+t =n_{i\bj},  \\
\text{even $i$ and odd $\bj$}&: \  - r_{i-1}+  \frac{  s_{{\overline{\jmath - 1}}}}{2}   +t =n_{i\bj},  \\
\text{even $i$ and even $\bj$}&: \ - r_{i-1} -  \frac{ s_{\bj} }{2}+t =n_{i\bj}. 
\end{cases}
\label{sln-period2}
\end{align}

Finally we have to  impose the bound (\ref{bound}) on the  gauge connection for the 
conical defect. 
For the $sl(N|N-1)$ case,   the charge $L_0$ in terms of the holonomy of the
background is given by 
\begin{equation} \label{vall-n}
L_0 = \frac{c}{24 \epsilon_{(N|N-1)}} \left( {\rm str} ( a_{\phi}^2)  \right) , 
\end{equation}
 and 
 \begin{align}\label{defep}
 \epsilon_{\left(N|N-1\right)} &= \text{str}(\hat L_0 \hat  L_0)=\frac{1}{4}N(N-1) . 
 \end{align}
 Here we have  used  the explicit representation of $\hat L_0$ given in (\ref{hatl}). 
Then equation (\ref{bound}) reduces to 
\begin{equation}
0< \sum_k a_k ^2 - \sum _{\bl } a_\bl ^2 + 4 N(N-1) \xi ^2 < \frac{N(N-1)}{8}. 
\end{equation}
 Now  substituting  the quantization conditions given  in  (\ref{holo-int}) for even and odd $N$ 
 the bound can be written as 
 \begin{align} \label{energy1}
\text{For even $N$} \ & :\ \ 0 < \frac{1}{4}  \underset{j \, \text{odd}}{ \sum_{j=1}^{N-1}} m_j^2 - \underset{j \, \text{odd}} {\sum_{\bj =\overline{ N+1}} ^{\overline{2N-3}}} p_{\bj}^2 - N(N-1)q^2 < \frac{N(N-1)}{8},  \\ \nn
\text{For odd $N$} \ & :\ \ 0 <  \underset{j \, \text{odd}}{ \sum_{j=1}^{N-1}} r_j^2 - \frac{1}{4}  \underset{j \, \text{even}} {\sum_{\bj =\overline{ N+1}} ^{\overline{2N-2}}} s_{\bj}^2 - N(N-1)t^2 < \frac{N(N-1)}{8} . 
 \end{align}
 The theory admits 
smooth as well as supersymmetric conical defects if the above bound together 
with the periodicity  conditions in (\ref{sln-period}) and (\ref{sln-period2}) are satisfied. 
We will now provide some simple examples to demonstrate that smooth supersymmetric
conical defects are allowed for $N\geq 5$. 
For the $N=5$ case  the  bound reduces to 
\begin{equation}
0< (r_1^2 +r_3^2) - \frac{1}{4} (s_{\bar{6}}^2 +s_{\bar{8}}^2 ) -20t^2< \frac{5}{2}.  
\end{equation}
This inequality is satisfied for $r_1= r_3=1, \ s_{\bar{6}} = s_{\bar{6}}=2, \ t=0$. While for $N=6$ we have
\begin{equation}
0< \frac{1}{4}(m_1^2 + m_3^2 + m_5^2) - (p_{\bar{7}}^2 + p_{\bar{9}}^2)-30q^2 <\frac{15}{4}. 
\end{equation}
This inequality is satisfied for $ m_1=m_2=m_3=2,\ p_{\bar{7}}=p_{\bar{9}}=1 , \ q=0  $.
It is clear from these examples that as $N$ gets larger the term on the extreme RHS of the 
inequalities in (\ref{energy1}) increases and it should be possible to find more integers 
$m_j, p_{\bar j}, r_j, s_{\bar j}, q, t$
to satisfy the inequality.

\subsubsection*{Smoothness and supersymmetry:  $sl(4|3)$ }

As mentioned earlier 
the case for $N=4$ needs to be treated separately since the form obtained 
by diagonalizing the connection given in (\ref{sln-gc})  can not be written 
in the general form  given in (\ref{lincomb}). 
Diagonalizing the background connection for $N=4$  and writing it as a linear 
combination of the Cartan generators we obtain
\begin{align}\label{hol43}
Sa_\phi S^{-1} = ia_1 H_1 +i a_3 H_3 + i a_{\bar 5 } (H_{\bar 5 } + H_{\bar 6} ). 
\end{align}
Imposing trivial holonomies for smoothness of the conical defect solutions, we get the 
following quantization conditions
\begin{align} \label{int43}
 a_i=\frac{m_i}{2}, \ a_{\bar{5}}= p_{\bar{5}} \ , \ -2i\xi =q . 
\end{align}
\def\bfive{{\overline{5}}}
\def\bsix{{\overline{6}}}
Here $p_{i},q , m_i\in \mathbb{Z} $ and $m_i$  takes values in either the set of 
even or odd integers for
all $i$. 
The commutator of \( Sa_\phi S^{-1} \) with fermionic generators $E_{i\bj}$ are as follows
\begin{align}
i=1,3 \ &  \begin{cases}
[ Sa_\phi S^{-1} , E_{i\bfive} ] = i(a_i-a_\bfive) E_{i\bfive},  \\
[ Sa_\phi S^{-1} , E_{i\bsix} ] = i a_i E_{i\bsix}, 
\end{cases} \\
i=2,4 \ & \begin{cases}
[ Sa_\phi S^{-1} , E_{i\bfive} ] = -i(a_{i-1} + a_\bfive) E_{i\bfive},  \\
[ Sa_\phi S^{-1} , E_{i\bsix} ] = - ia_{i-1} E_{i\bsix}. 
\end{cases}
\end{align}
Substituting these roots in the supersymmetry conditions (\ref{periodf}) we obtain
 the following periodicity conditions
\begin{align}\label{per43}
&i(a_1 - a_\bfive) + 2\xi =in_{1\bfive}, \quad ia_1 + 2\xi =in _{1\bsix},  \quad i (a_3 - a_\bfive) + 2\xi =in_{3\bfive}, \quad i a_3 + 2\xi =in _{3\bsix},\nn  \\
-&i(a_1 + a_\bfive) + 2\xi =in_{2\bfive}, \  -ia_1 + 2\xi =in _{2\bsix}, \ -i(a_3 + a_\bfive) + 2\xi =in_{4\bfive}, \   -ia_3 + 2\xi =in _{4\bsix}.
\end{align}
The $a_i$, $a_{\bi}$ and $\xi$ are further constrained by (\ref{int43}). 

Upon imposing the energy bound condition  (\ref{bound})
and using  the quantization conditions in (\ref{int43}) we obtain the following inequality
\begin{equation}
0< \frac{1}{4}(m_1^2 + m_3^2) - p_\bfive ^2 - 12 q^2 < \frac{3}{2}, 
\end{equation}
A simple example in which this inequality is satisfied along with the constraints in (\ref{per43})
is
\begin{equation}
m_1 =2, \quad m_3 =p_{\bar 5} = q=0. 
\end{equation}
Thus  smooth supersymmetric conical defects therefore do exist in the $sl(4|3)$ theory.

\section{Conclusions}

The main result of this paper is the observation that the  supersymmetry conditions
of a given background for the $sl(N|N-1)$  theory can be written down in terms
of products of the eigenvalues of the background holonomies with the odd roots of the 
super algebra.  
Thus the  periodicity constraint on  the Killing spinor can be formulated 
in terms of gauge invariant and physically independent observables. 
This condition is given in (\ref{periodf}).
We have also constructed a class of conical defects and black holes in the $sl(3|2)$ theory. 
These solutions in general have fields in $sl(3)\oplus sl(2) \oplus u(1)$ directions turned on.
We have obtained the periodicity properties for  
 the Killing spinors  in these backgrounds explicitly 
 by solving the Killing spinor equations.  A summary of the solutions and the supersymmetry
 conditions is given in table 1 of \ref{table}. 
 These conditions can be  seen to be in agreement with the general constraint (\ref{periodf}).
 Though the analysis which resulted in 
 the supersymmetry condition given in (\ref{periodf}) was done in the $sl(N|N-1)$  as 
 a concrete example,
 the condition (\ref{periodf}) is general  and  can be applied to  a 
Chern-Simons theories based on any super group. 

We have shown that for $N\geq 4$, the $sl(N|N-1)$ admits smooth supersymmetric
conical defects. Just as smooth conical defects in the bosonic $sl(N)$ theory are
dual to the primaries of the ${\cal W}_N$ minimal model, the smooth supersymmetric
conical defects should be dual to the chiral primaries of the supersymmetric 
minimal model conjectured to be the  large $N$ limit
of the the $sl(N|N-1)$ theories \cite{Creutzig:2011fe}.  
It will be interesting to classify the chiral primaries 
of these supersymmetric minimal models  and compare them with 
the supersymmetric conical defects found in this paper \cite{shouvik}. 
Conical surplus solutions
 in the bosonic $sl(N,C)$ Chern-Simons  
theory have been shown to agree with the light states of the dual minimal model 
\cite{Castro:2011iw}. 
It will be interesting to see if  the Euclidean supersymmetric version of the Chern-Simons
theory studied in this paper 
admits conical surplus solutions and check if they are supersymmetric. 
One can then verify if they correspond to possible light states in the 
dual Kazama-Suzuki model of \cite{Creutzig:2011fe}\footnote{
We thank Rajesh Gopakumar for discussions regarding this point}. 

The black holes we constructed in the $sl(3|2)$ theory have in addition to 
fields in $sl(3)$ also fields in the extra  $sl(2)$ turned on. 
It will be interesting to study their smoothness/holonomy  and 
 the thermodynamic properties of these black hole solutions
and obtain an expression for their partition function both from the bulk theory and the 
CFT.

\acknowledgments

We wish to thank Rajesh Gopakumar and S. Prem Kumar for useful and stimulating discussions. 
The work of J.R.D is partially supported by the Ramanujan fellowship DST-SR/S2/RJN-59/2009.

\appendix
\section{Cartan-Weyl basis for $sl(N|N-1)$ }
\def\beo{\begin{eqnarray}}
\def\eno{\end{eqnarray}}
\def\bk{{\bar k}}
One can construct following {\cite{Frappat:1996pb}} a basis of matrices for the $sl(N|N-1)$ algebra. Let's consider $(2N-1)^2$ matrices $e_{IJ}$ of order $2N-1$ so that $(e_{IJ})_{KL} = \delta_{IL} \delta_{JK}$ ($I,J,K,L = 1,\dots,2N-1$) and define the 
$(2N-1)^2-1$ generators
\beo
&&E_{ij} = e_{ij} -   \delta_{ij} (e_{kk}+e_{\bk\bk}) , 
~~~~~~~~~~ E_{i \bj} = e_{i \bj},  \\
&&E_{\bi\bj} = e_{\bi\bj} +   \delta_{\bi\bj}
(e_{kk}+e_{\bk\bk}) , 
~~~~~~~~~~ E_{\bi j} = e_{\bi j}, 
\eno
  where $i,j,\dots$ run from 1 to $N$ and $\bi,\bj,\dots$ from 
$N+1$ to $2N-1$.\\ 
The generators for the various subalgebras of $sl(3|2)$ are as follows
\begin{itemize}
\item  $u(1)$ : $J = E_{kk} = -E_{\bk\bk} =
- ((N-1) e_{kk} + N e_{\bk\bk})$. \\
 For the above mentioned matrix representation we get $$J = (-(N-1))\mathbf{1}_{N\times N} \oplus (-N) \mathbf{1}_{(N-1)\times (N-1)}$$ It then follows that $\text{str}(J^2)=-N(N-1)$. 
\item $sl(N)$:  $E_{ij} - \tfrac{1}{N} \delta_{ij} Z$. 

\item $sl(N-1)$  : $E_{\bi\bj} + \tfrac{1}{N-1}
\delta_{\bi\bj} Z$. 

\item $(\overline{N},N-1)$ representation of $sl(N) \oplus sl(N-1) \oplus u(1)$ : $E_{i\bj}$. 

\item $(N,\overline{N-1})$
representation of $sl(N) \oplus sl(N-1) \oplus u(1)$ : $E_{\bi j}$. 

\end{itemize}\vspace{0.3cm}
In the Cartan-Weyl basis, the generators are given by
\begin{itemize}
\item Cartan subalgebra
\begin{align}
&H_i = E_{ii} - E_{i+1,i+1},  \quad  \text{for} \quad   1\leq i \leq N-1,  \\
&H_{\bi} = E_{\bi\bi} - E_{\bi +1,\bi +1},  \quad \text{for} \quad  N+1\leq i \leq 2N-2,  \\
&H_N = E_{NN} +E_{N+1,N+1}. 
\end{align}

\item Raising operators
\begin{align}
E_{ij} \ \text{with $i<j$ for $sl(N)$}, \quad E_{\bi\bj} \ \text{with $\bi<\bj$ for $sl(N-1)$}, \quad E_{i\bj}.  \ \text{for the odd part}
\end{align}

\item Lowering operators
\begin{align}
E_{ji} \ \text{with $i<j$ for $sl(N)$}, \quad E_{\bj\bi} \ \text{with $\bi<\bj$ for $sl(N-1)$}, \quad E_{\bi j}.  \ \text{for the odd part}
\end{align}

\end{itemize}

\def\cle{\left[}
\def\cri{\right]}
\def\ale{\left\lbrace}
\def\ari{\right\rbrace}
The commutation relations in this basis are
\beo
&&\cle H_I,H_J \cri = 0,  \nn \\
&&\cle H_K,E_{IJ} \cri = \delta_{IK} E_{KJ} - \delta_{I,K+1} E_{K+1,J} 
 - \delta_{KJ} E_{IK} + \delta_{K+1,J} E_{I,K+1} 
 \qquad (K \ne N),  \nn \\ 
&&\cle H_N,E_{IJ} \cri = \delta_{Im} E_{NJ} + \delta_{I,N+1} E_{N+1,J} 
 - \delta_{NJ} E_{Im} - \delta_{N+1,J} E_{I,N+1}, \nn \\
&&\cle E_{IJ},E_{KL} \cri = \delta_{JK} E_{IL} - \delta_{IL} E_{KJ} 
\qquad \text{for $E_{IJ}$ and $E_{KL} $  even},  \\
&&\cle E_{IJ},E_{KL} \cri = \delta_{JK} E_{IL} - \delta_{IL} E_{KJ}
\qquad \text{ for $E_{IJ}$   even and $E_{KL}$  odd}, \nn \\
&&\ale E_{IJ},E_{KL}\ari = \delta_{JK} E_{IL} + \delta_{IL} E_{KJ}
\qquad  \text{  for $E_{IJ}$ and $E_{KL}$  odd}.  \nn
\eno

\providecommand{\href}[2]{#2}\begingroup\raggedright\endgroup


\begin{thebibliography}{10}

\bibitem{Vasiliev:2003ev}
M.~Vasiliev, {\it {Nonlinear equations for symmetric massless higher spin
  fields in (A)dS(d)}},  {\em Phys.Lett.} {\bf B567} (2003) 139--151,
  [\href{http://xxx.lanl.gov/abs/hep-th/0304049}{{\tt hep-th/0304049}}].

\bibitem{Bekaert:2005vh}
X.~Bekaert, S.~Cnockaert, C.~Iazeolla, and M.~Vasiliev, {\it {Nonlinear higher
  spin theories in various dimensions}},
  \href{http://xxx.lanl.gov/abs/hep-th/0503128}{{\tt hep-th/0503128}}.

\bibitem{Klebanov:2002ja}
I.~Klebanov and A.~Polyakov, {\it {AdS dual of the critical O(N) vector
  model}},  {\em Phys.Lett.} {\bf B550} (2002) 213--219,
  [\href{http://xxx.lanl.gov/abs/hep-th/0210114}{{\tt hep-th/0210114}}].

\bibitem{Giombi:2009wh}
S.~Giombi and X.~Yin, {\it {Higher Spin Gauge Theory and Holography: The
  Three-Point Functions}},  {\em JHEP} {\bf 1009} (2010) 115,
  [\href{http://xxx.lanl.gov/abs/0912.3462}{{\tt arXiv:0912.3462}}].

\bibitem{Giombi:2010vg}
S.~Giombi and X.~Yin, {\it {Higher Spins in AdS and Twistorial Holography}},
  {\em JHEP} {\bf 1104} (2011) 086,
  [\href{http://xxx.lanl.gov/abs/1004.3736}{{\tt arXiv:1004.3736}}].

\bibitem{Sundborg:2000wp}
B.~Sundborg, {\it {Stringy gravity, interacting tensionless strings and
  massless higher spins}},  {\em Nucl.Phys.Proc.Suppl.} {\bf 102} (2001)
  113--119, [\href{http://xxx.lanl.gov/abs/hep-th/0103247}{{\tt
  hep-th/0103247}}].

\bibitem{Mikhailov:2002bp}
A.~Mikhailov, {\it {Notes on higher spin symmetries}},
  \href{http://xxx.lanl.gov/abs/hep-th/0201019}{{\tt hep-th/0201019}}.

\bibitem{Sezgin:2002rt}
E.~Sezgin and P.~Sundell, {\it {Massless higher spins and holography}},  {\em
  Nucl.Phys.} {\bf B644} (2002) 303--370,
  [\href{http://xxx.lanl.gov/abs/hep-th/0205131}{{\tt hep-th/0205131}}].

\bibitem{Douglas:2010rc}
M.~R. Douglas, L.~Mazzucato, and S.~S. Razamat, {\it {Holographic dual of free
  field theory}},  {\em Phys.Rev.} {\bf D83} (2011) 071701,
  [\href{http://xxx.lanl.gov/abs/1011.4926}{{\tt arXiv:1011.4926}}].

\bibitem{Chang:2012kt}
C.-M. Chang, S.~Minwalla, T.~Sharma, and X.~Yin, {\it {ABJ Triality: from
  Higher Spin Fields to Strings}},
  \href{http://xxx.lanl.gov/abs/1207.4485}{{\tt arXiv:1207.4485}}.

\bibitem{Blencowe:1988gj}
M.~Blencowe, {\it {A consistent interacting masless higher spin field theory in
  D = (2+1)}},  {\em Class.Quant.Grav.} {\bf 6} (1989) 443.

\bibitem{Gaberdiel:2010pz} 
  M.~R.~Gaberdiel and R.~Gopakumar,
  {\it {An $AdS_3$ Dual for Minimal Model CFTs}},
  {\em Phys.Rev.} {\bf D83}  (2011) 066007, 
 [\href{http://xxx.lanl.gov/abs/1011.2986}{{\tt arXiv:1011.2986}}].


\bibitem{Gaberdiel:2012uj}
M.~R. Gaberdiel and R.~Gopakumar, {\it {Minimal Model Holography}},
  \href{http://xxx.lanl.gov/abs/1207.6697}{{\tt arXiv:1207.6697}}.

\bibitem{Creutzig:2011fe}
T.~Creutzig, Y.~Hikida, and P.~B. Ronne, {\it {Higher spin {$ AdS_3 $}
  supergravity and its dual CFT}},  {\em JHEP} {\bf 1202} (2012) 109,
  [\href{http://xxx.lanl.gov/abs/1111.2139}{{\tt arXiv:1111.2139}}].

\bibitem{Candu:2012jq}
C.~Candu and M.~R. Gaberdiel, {\it {Supersymmetric holography on $AdS_3$}},
  \href{http://xxx.lanl.gov/abs/1203.1939}{{\tt arXiv:1203.1939}}.

\bibitem{Candu:2012tr}
C.~Candu and M.~R. Gaberdiel, {\it {Duality in N=2 minimal model holography}},
  \href{http://xxx.lanl.gov/abs/1207.6646}{{\tt arXiv:1207.6646}}.

\bibitem{Henneaux:2012ny}
M.~Henneaux, G.~Lucena~Gomez, J.~Park, and S.-J. Rey, {\it {Super- W(infinity)
  Asymptotic Symmetry of Higher-Spin $AdS_3$ Supergravity}},  {\em JHEP} {\bf
  1206} (2012) 037, [\href{http://xxx.lanl.gov/abs/1203.5152}{{\tt
  arXiv:1203.5152}}].

\bibitem{Gutperle:2011kf}
M.~Gutperle and P.~Kraus, {\it {Higher Spin Black Holes}},  {\em JHEP} {\bf
  1105} (2011) 022, [\href{http://xxx.lanl.gov/abs/1103.4304}{{\tt
  arXiv:1103.4304}}].

\bibitem{Castro:2011iw}
A.~Castro, R.~Gopakumar, M.~Gutperle, and J.~Raeymaekers, {\it {Conical Defects
  in Higher Spin Theories}},  {\em JHEP} {\bf 1202} (2012) 096,
  [\href{http://xxx.lanl.gov/abs/1111.3381}{{\tt arXiv:1111.3381}}].

\bibitem{Ammon:2011nk}
M.~Ammon, M.~Gutperle, P.~Kraus, and E.~Perlmutter, {\it {Spacetime Geometry in
  Higher Spin Gravity}},  {\em JHEP} {\bf 1110} (2011) 053,
  [\href{http://xxx.lanl.gov/abs/1106.4788}{{\tt arXiv:1106.4788}}].

\bibitem{Castro:2011fm}
A.~Castro, E.~Hijano, A.~Lepage-Jutier, and A.~Maloney, {\it {Black Holes and
  Singularity Resolution in Higher Spin Gravity}},  {\em JHEP} {\bf 1201}
  (2012) 031, [\href{http://xxx.lanl.gov/abs/1110.4117}{{\tt
  arXiv:1110.4117}}].

\bibitem{Didenko:2009td}
  V.~E.~Didenko and M.~A.~Vasiliev,
  {\it {Static BPS black hole in 4d higher-spin gauge theory}}, 
  {\em Phys.\ Lett.\ B } {\bf 682} (2009) 305, 
  [\href{http://xxx.lanl.gov/abs/0906.3898}{{\tt
  arXiv:0906.3898}}].

    
\bibitem{Coussaert:1993jp}
O.~Coussaert and M.~Henneaux, {\it {Supersymmetry of the (2+1) black holes}},
  {\em Phys.Rev.Lett.} {\bf 72} (1994) 183--186,
  [\href{http://xxx.lanl.gov/abs/hep-th/9310194}{{\tt hep-th/9310194}}].

\bibitem{Izquierdo:1994jz}
J.~Izquierdo and P.~Townsend, {\it {Supersymmetric space-times in (2+1) adS
  supergravity models}},  {\em Class.Quant.Grav.} {\bf 12} (1995) 895--924,
  [\href{http://xxx.lanl.gov/abs/gr-qc/9501018}{{\tt gr-qc/9501018}}].

\bibitem{David:1999zb}
J.~R. David, G.~Mandal, S.~Vaidya, and S.~R. Wadia, {\it {Point mass
  geometries, spectral flow and AdS(3) - CFT(2) correspondence}},  {\em
  Nucl.Phys.} {\bf B564} (2000) 128--141,
  [\href{http://xxx.lanl.gov/abs/hep-th/9906112}{{\tt hep-th/9906112}}].

\bibitem{Balasubramanian:2000rt}
V.~Balasubramanian, J.~de~Boer, E.~Keski-Vakkuri, and S.~F. Ross, {\it
  {Supersymmetric conical defects: Towards a string theoretic description of
  black hole formation}},  {\em Phys.Rev.} {\bf D64} (2001) 064011,
  [\href{http://xxx.lanl.gov/abs/hep-th/0011217}{{\tt hep-th/0011217}}].

\bibitem{Maldacena:2000dr}
J.~M. Maldacena and L.~Maoz, {\it {Desingularization by rotation}},  {\em JHEP}
  {\bf 0212} (2002) 055, [\href{http://xxx.lanl.gov/abs/hep-th/0012025}{{\tt
  hep-th/0012025}}].

\bibitem{Romans:1991wi}
L.~Romans, {\it {The N=2 super W(3) algebra}},  {\em Nucl.Phys.} {\bf B369}
  (1992) 403--432.

\bibitem{Tan:2012xi}
H.~Tan, {\it {Exploring Three-dimensional Higher-Spin Supergravity based on
  {$sl(N |N - 1)$} Chern-Simons theories}},
  \href{http://xxx.lanl.gov/abs/1208.2277}{{\tt arXiv:1208.2277}}.

\bibitem{Witten:1988hc}
E.~Witten, {\it {(2+1)-Dimensional Gravity as an Exactly Soluble System}},
  {\em Nucl.Phys.} {\bf B311} (1988) 46.

\bibitem{Achucarro:1987vz}
A.~Achucarro and P.~Townsend, {\it {A Chern-Simons Action for Three-Dimensional
  anti-De Sitter Supergravity Theories}},  {\em Phys.Lett.} {\bf B180} (1986)
  89.

\bibitem{Henneaux:2010xg} 
  M.~Henneaux and S.~-J.~Rey,
  {\it {Nonlinear $W_{infinity}$ as Asymptotic Symmetry of Three-Dimensional Higher Spin Anti-de Sitter Gravity}}, 
 {\em  JHEP} {\bf 1012}, (2010)  007, 
 [\href{http://xxx.lanl.gov/abs/1008.4579}{{\tt arXiv:1008.4579}}].

\bibitem{Campoleoni:2010zq}
A.~Campoleoni, S.~Fredenhagen, S.~Pfenninger, and S.~Theisen, {\it {Asymptotic
  symmetries of three-dimensional gravity coupled to higher-spin fields}},
  {\em JHEP} {\bf 1011} (2010) 007,
  [\href{http://xxx.lanl.gov/abs/1008.4744}{{\tt arXiv:1008.4744}}].


\bibitem{Frappat:1996pb}
L.~Frappat, P.~Sorba, and A.~Sciarrino, {\it {Dictionary on Lie
  superalgebras}},  \href{http://xxx.lanl.gov/abs/hep-th/9607161}{{\tt
  hep-th/9607161}}.

\bibitem{gomez}
{Gomez-Munoz, J. L, et. al.}, {\it {Quantum Add-on v2.0.3 for 
Mathematica}}, (2011),
  [\href{http://homepage.cem.itesm.mx/lgomez/quantum/}{{\tt 
http://homepage.cem.itesm.mx/lgomez/quantum/}}]. 



\bibitem{Henneaux:1999ib}
M.~Henneaux, L.~Maoz, and A.~Schwimmer, {\it {Asymptotic dynamics and
  asymptotic symmetries of three-dimensional extended AdS supergravity}},  {\em
  Annals Phys.} {\bf 282} (2000) 31--66,
  [\href{http://xxx.lanl.gov/abs/hep-th/9910013}{{\tt hep-th/9910013}}].

\bibitem{shouvik}
S.~Datta and J.~David, {\it {Work in progress}}.

\end{thebibliography}

\end{document}